\documentclass[aps,twocolumn,10pt,superscriptaddress,prb]{revtex4-1}

\usepackage{amsmath,amssymb,amsfonts}
\usepackage{graphicx}
\usepackage{braket}
\usepackage{nicefrac}
\usepackage{multirow}
\usepackage{cellspace}
\usepackage[colorlinks=True,linkcolor=red,citecolor=blue,urlcolor=blue]{hyperref}
\usepackage{bbold}

\newcommand{\me}{\mathrm{e}}
\newcommand{\mi}{\mathrm{i}}
\newcommand{\parent}[1]{\left(#1\right)}

\newcommand{\dd}{\mathrm{d}}
\newcommand{\vf}{v_{\mathrm{F}}}

\newcommand{\vmod}[1]{\left|#1\right|}

\begin{document}

\title{Linear optical conductivity of chiral multifold fermions}

\author{Miguel-\'Angel S\'anchez-Mart\'inez}
\affiliation{Univ. Grenoble Alpes, CNRS, Grenoble INP, Institut N\'eel, 38000 Grenoble, France}
\author{Fernando de Juan}
\affiliation{Donostia International Physics Center, 20018 Donostia-San Sebastian, Spain}
\affiliation{IKERBASQUE, Basque Foundation for Science, Maria Diaz de Haro 3, 48013 Bilbao, Spain}
\author{Adolfo G. Grushin}
\affiliation{Univ. Grenoble Alpes, CNRS, Grenoble INP, Institut N\'eel, 38000 Grenoble, France}

\begin{abstract}
Chiral multifold fermions are quasiparticles described by higher spin generalizations of the Weyl equation, and are realized as low energy excitations near symmetry protected band crossings in certain chiral crystals. In this work we calculate the linear optical conductivity of all chiral multifold fermions. We show that it is enhanced with respect to that of Weyl fermions with the same Fermi velocity, and features characteristic activation frequencies for each multifold fermion class, providing an experimental fingerprint to detect them. 
We calculate the conductivity for realistic chiral multifold semimetals by using lattice tight-binding Hamiltonians
that match the effective models of multifold fermions at low energies, for space groups 199 and 198. The latter includes RhSi, for which we give quantitative predictions, and also CoSi and AlPt. Our predictions can be tested in absorption or penetration depth measurements, and are necessary to extract the recently proposed quantized photocurrents from experiments.
\end{abstract}

\maketitle

\section{Introduction}
One of the clearest differences between topological metals and other metals is their electronic response to light. In TaAs, a prototypical Weyl semimetal, the bands disperse linearly from a protected twofold band crossing point, known as the Weyl node~\cite{armitage_weyl_2018,Gao_Heng}. Because of the absence of an energy scale, the linear optical conductivity is proportional to the driving frequency $\omega$~\cite{armitage_weyl_2018,Andolina,BurkovBalents2011,hosur_charge_2012,ashby_chiral_2014,tabert_optical_2016,Jenkins2016,mukherjee_doping_2018,Hutt,Kimura2017,Neubauer2018,crassee_3d_2018}, differing from that of systems with quadratically dispersing bands.

The absence of inversion symmetry, a common property to most known Weyl semimetals, allows a finite non-linear optical current proportional to even powers of the electric field. Most notably, second order photocurrents, that are proportional to the intensity of the electric field, have been predicted~\cite{Chan2016chiralphotons,YangBurch2017,de_juan_quantized_2017} and measured to be large in Weyl semimetals~\cite{Burch17,Ma:ux,Sun17,Wu17,Patankar2018,Sirica2018,GaoQin2019}. For example, second harmonic generation, a current oscillating at twice the frequency of the incident light, has record breaking magnitudes in the monopnictide TaAs class of topological semimetals~\cite{Wu17}, resonantly enhanced at low frequencies~\cite{Patankar2018}. Additionally, semimetals that not only break inversion symmetry but also all mirror symmetries~\cite{Chang:2018bb} are expected to generate a large and quantized non-linear photocurrent induced by circularly polarized light~\cite{de_juan_quantized_2017}.

Less is known about the optical responses of the recent members in the family of topological metals, known as multifold semimetals~\cite{manes_existence_2012,bradlyn_beyond_2016,tang_multiple_2017}. Multifold semimetals are characterized by protected band crossings of degeneracy higher than two, and generalize the concept of Weyl semimetals. The quasiparticles at energies close to these crossing points, called multifold fermions, are governed by Weyl-like Hamiltonians: pseudo-relativistic and linear in momentum and effective spin, of the form $H=\hbar v_F\mathbf{k}\cdot \mathbf{S}$. They exist as either three, four, six or eight fold degeneracies, of which only the first three can be chiral. This means that only the first three types can have bands characterized by a topological invariant, the Chern number, defining the multifold crossings as monopoles of Berry flux. 

Multifold fermions are the most promising candidates to display a quantized circular photogalvanic effect~\cite{chang_unconventional_2017,flicker_chiral_2018}.
Experiments using angle resolved photoemission spectroscopy (ARPES) in CoSi~\cite{takane_observation_2018,Rao2018,sanchez_discovery_2018}, AlPt~\cite{schroter_topological_nodate} and RhSi~\cite{sanchez_discovery_2018}, all in space group (SG) 198, are consistent with the existence of chiral multifold fermions at the Fermi energy in these materials~\cite{bradlyn_beyond_2016,tang_multiple_2017,Pshenay_Severin_2018}. Additionally, a frequency independent photovoltaic plateau was detected in RhSi~\cite{Rees:2019ue}, consistent with the expected photogalvanic quantization~\cite{de_juan_quantized_2017,chang_unconventional_2017,flicker_chiral_2018}. However, to faithfully extract the quantized non-linear conductivity, and to further confirm that multifold fermions are the low energy quasiparticles in these materials, a good knowledge of the absorption, determined by the linear optical conductivity, is needed~\cite{de_juan_quantized_2017,Rees:2019ue}, yet currently absent. 

In this work we calculate the linear optical conductivity, defined as the linear response coefficient relating the applied electric field to the induced current, for all chiral multifold fermions. We describe how they can be distinguished by this observable, and provide predictions for real materials. We find that all types of chiral multifolds have an optical conductivity larger than a Weyl semimetal with the same Fermi velocity $v_F$. Moreover, the frequencies at which different allowed transitions are activated distinguish each multifold fermion. We therefore find that the optical conductivity provides a clear fingerprint of each chiral multifold fermion, similar to their two dimensional counterparts~\cite{Dora}. We use this knowledge to predict the linear optical conductivity of materials in space group (SG)198 and SG 199. Specifically, we calculate the linear optical conductivity of RhSi, which determines its reflection and absorption and can be measured by ellipsometry.

The paper is structured as follows. In Sec.~\ref{sec:lowenergymodels} we provide the general formulas used and their connection to experimental measurements, discussing first low energy models without spin-orbit coupling that we then generalize to include spin-orbit coupling. In Sec.~\ref{sec:realistic} we use realistic tight-binding models to predict the linear optical conductivity of RhSi, as well as for materials in SG199. Finally, in Sec.~\ref{sec:conclusions} we summarize and discuss our results. An explicit calculation of the imaginary part of the optical conductivity using Kramers-Kronig relations, the sum rules associated to the longitudinal conductivity and additional details of our calculation are provided in the appendices.

\section{Optical Conductivity}

The conductivity $\sigma_{\mu\nu}$  of a material is the linear response coefficient between an electric field applied in the $\nu$ direction and the current density induced in the $\mu$ direction. If the applied electric field has a wavelength larger than the lattice constant, the momentum $\mathbf{q}$ transferred by the photon to the electron is negligible and the electron conserves its momentum $\mathbf{k}$ in the process. We refer to the conductivity in this limit $\mathbf{q}\to 0$ as optical conductivity $\sigma_{\mu\nu}(\omega)$, which depends on the electric field's frequency $\omega$.  When $\omega$ is sufficiently large to overcome Pauli blocking, an incident photon excites one electron from an occupied state to an unoccupied state. This process, known as an interband transition contribution to the optical conductivity, can be calculated using standard linear response theory as the real part of~\cite{Mahan}

\begin{align}
\label{eqn:optcond}
\sigma_{\mu \nu}(\omega)=\frac{ie^2}{\omega V}\sum_{m\neq n}\frac{\bra{n} j_{\mu}\ket{m}\bra{m}j_{\nu}\ket{n}}{\epsilon_n-\epsilon_m+\hbar\omega+i\delta}\left(n_F(\epsilon_{n})-n_F(\epsilon_{m})\right),
\end{align}
where $e$ is the charge of the electron, $j_{\mu}=\frac{1}{\hbar}\partial_{k_{\mu}}H$ is the current operator associated with the Hamiltonian $H$ describing the system, $V$ is the volume of the sample, $\ket{n}$ and $E_n$ are an eigenstate of $H$ and its corresponding eigenvalue, respectively, $\epsilon_n=E_n-\mu$ with $\mu$ the chemical potential, and $\delta$ is an infinitesimal broadening. The Fermi function $n_F$ depends on $\epsilon_n$, $\mu$ and the inverse temperature $\beta=1/k_B T$ measured in units of the Boltzmann constant $k_B$. 

Our goal is to calculate the interband contribution to the optical conductivity (Eq.~\ref{eqn:optcond}) of all chiral multifold fermions. Since these occur in cubic space groups, the three diagonal elements $\sigma_{xx}$, $\sigma_{yy}$, and $\sigma_{zz}$ are equal and we can focus on a single component, $\sigma_{xx}$~\footnote{There is a type of double spin-1/2 fermion that can occur in non-cubic space groups. Since our results will not change qualitatively and RhSi, CoSi and AlPt are cubic, we restrict our analysis to cubic space groups.}.
In the main body of this work we will compute the real part of the interband optical conductivity, and obtain its imaginary part using standard Kramers-Kronig relations~\cite{dresselhaus_optical_nodate} in Appendix~\ref{sec:KK}.
There exists an additional Fermi surface contribution to the conductivity, the intraband Drude-like term, that scales as $1/\omega$ when $\omega\to 0$ and will be dominant at small frequencies. Since this contribution is not different from any other metal we omit it in the discussion that follows.

\section{Optical conductivity of Multifold Fermions: Low energy models}
\label{sec:lowenergymodels}
\subsection{Multifold fermions}
\label{subsec:lowenergysym}

Multifold fermions are low energy excitations that exist close to points in momentum space where linearly dispersing bands meet. The simplest example is the crossing of two bands, a Weyl fermion, which is protected against the opening of a gap so long as it is isolated in the Brillouin Zone. If more than two bands meet, the degeneracy point is not robust against perturbations that lift the degeneracy unless additional lattice symmetries protect it. 
Excitations around these protected crossings are called multifold fermions and can only exist as three-, four-, six- or eightfold  degeneracies. Due to their importance to non-linear optics and recent experimental realization we focus on chiral multifolds~\cite{manes_existence_2012,bradlyn_beyond_2016}: three-, four- and sixfold crossings. A pedagogical introduction to chiral multifold fermions, classified by Refs.~\onlinecite{manes_existence_2012,bradlyn_beyond_2016}, can be found in Ref.~\onlinecite{flicker_chiral_2018}.

The low-energy degrees of freedom near chiral multifold crossings of degeneracy larger than two can be described by a generalization of a Weyl Hamiltonian of the form $H=\hbar v_F\mathbf{k}\cdot \mathbf{S}_{\alpha}$, where $\mathbf{S}_{\alpha}$ is a vector of three matrices that depend on a material-specific parameter $\alpha$. For particular values $\alpha=\alpha_0$, only achieved without spin-orbit coupling, the matrices $\mathbf{S}_{\alpha}$ take the rotationally symmetric form of a higher-spin representation of SU(2). In such cases, the multifold fermions have an effective spin given by $\mathbf{S}_{\alpha_0}$. In the next subsections we calculate the optical conductivity for $\alpha=\alpha_0$, generalizing then to arbitrary values of $\alpha$.

To calculate the optical conductivity of all chiral multifold fermions it is helpful to note that, at linear order, some high-degeneracy multifolds can be decomposed into two decoupled Hamiltonians of lower degeneracy~\cite{flicker_chiral_2018}. This is the case for the sixfold fermion, which can be expressed as the direct sum of two threefold degeneracies: the Hamiltoninan describing a sixfold can always be brought to a block diagonal from, composed of two decoupled threefold Hamiltonians. Additionally, out of the two types of fourfold fermions that exist, only one can be written as a Hamiltonian consisting of two decoupled Weyl fermions of the same chirality~\cite{bradlyn_beyond_2016,flicker_chiral_2018}. We will refer to this case as a double spin-$1/2$ fourfold. The second type, which we will refer to as spin-$3/2$ fourfold fermion, cannot be expressed as the combination of lower degeneracy multifolds. Hence, it is enough to calculate the optical conductivity of a Weyl, a threefold, and a spin-$3/2$ fourfold fermion, since all chiral multifold fermions are built out of these three types.

\subsection{Optical conductivity in fully rotational symmetric models}
\label{subsec:optcondsym}

\begin{table}
  \includegraphics[width=1.\linewidth]{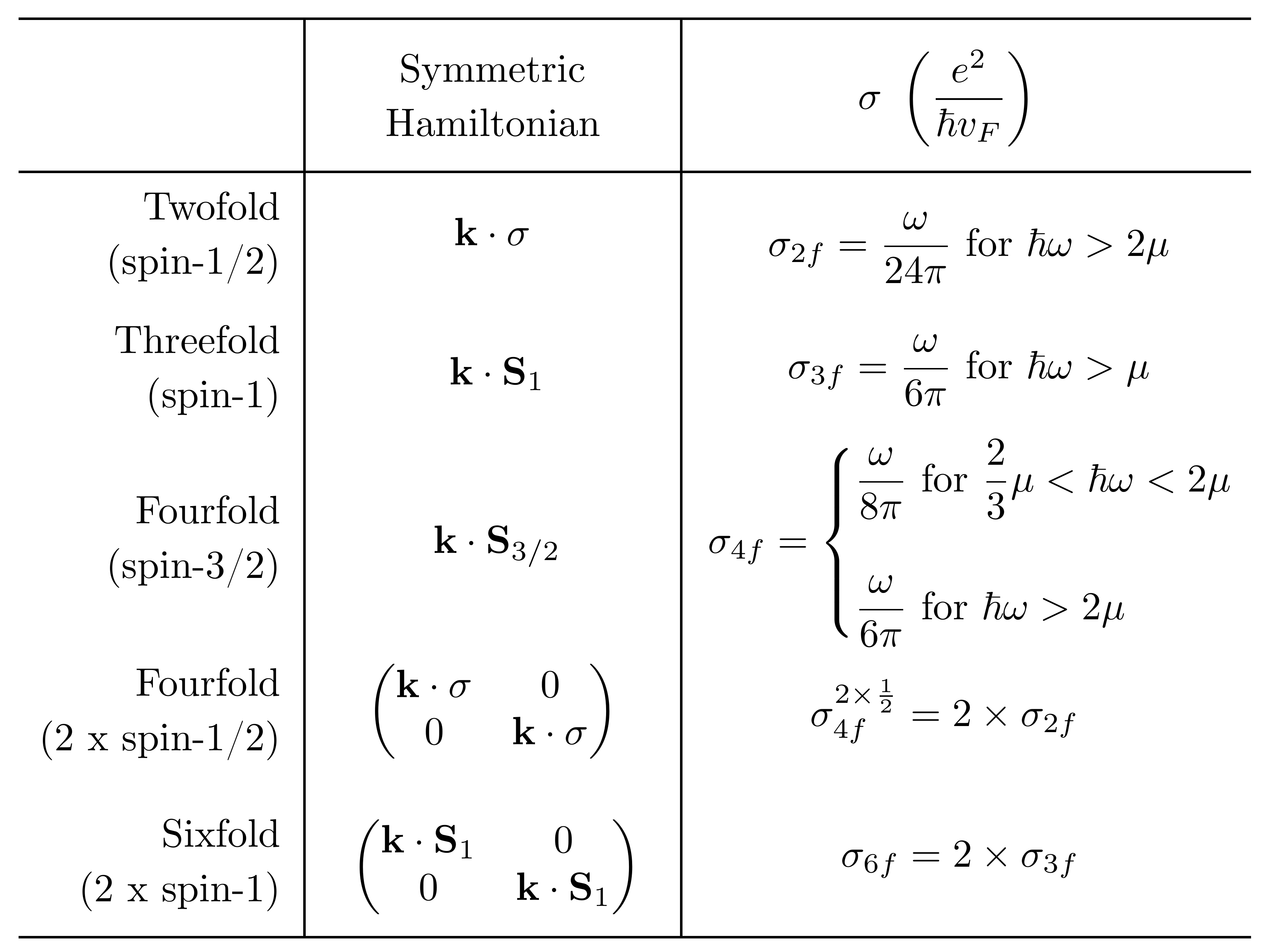}
  \caption{Effective Hamiltonians (in units of $1/\hbar v_F$) and their corresponding optical conductivities for all symmetric chiral multifold fermions. The optical conductivity of the effective models for the twofold, threefold and spin-$3/2$ fourfold fermions, discussed Sec.~\ref{subsec:lowenergysym}, are defined piecewise for each region delimited by their characteristic frequencies. The effective Hamiltonian of the double spin-$1/2$ fourfold is a direct sum of two Weyl Hamiltonians, and its optical conductivity is twice that of the Weyl fermion. Similarly, the effective Hamiltonian of the sixfold fermion is the direct sum of two threefold Hamiltonians, and its optical conductivity is two times that of the threefold fermion.}
  \label{table:optconds}
\end{table}

The lowest-degeneracy multifold fermion is the twofold, known as a Weyl fermion. The low-energy degrees of freedom near this twofold crossing are described by the Weyl Hamiltonian $H=\hbar v_F \mathbf{k}\cdot\boldsymbol{\sigma}$, where $\boldsymbol{\sigma}$ is a vector of Pauli matrices and $\mathbf{k}$ is the momentum. A simple dimensional analysis of Eq.~\eqref{eqn:optcond} using the Weyl Hamiltonian shows that the optical conductivity of Weyl fermions must have a linear dependence on the frequency $\omega$~\cite{BurkovBalents2011,hosur_charge_2012,ashby_chiral_2014}, and its explicit computation gives as a result~\cite{tabert_optical_2016}   
\begin{equation}
\label{eqn:condweyl}
\sigma_{W}(\omega)=\frac{\omega e^2}{24\pi\hbar v_F}\frac{\sinh(\hbar\omega\beta/2)}{\cosh(\mu\beta)+\cosh(\hbar\omega\beta/2)}.
\end{equation}
In the limit of zero temperature Eq.~\eqref{eqn:condweyl} takes the form~\cite{BurkovBalents2011,hosur_charge_2012,ashby_chiral_2014,tabert_optical_2016} $\sigma_{W}(\omega)=\frac{\omega e^2}{24\pi v_F\hbar}\Theta(\hbar\omega-2\mu)$, where $\Theta(x)$ is the Heaviside step function.

\begin{figure}
    \includegraphics[width=\linewidth]{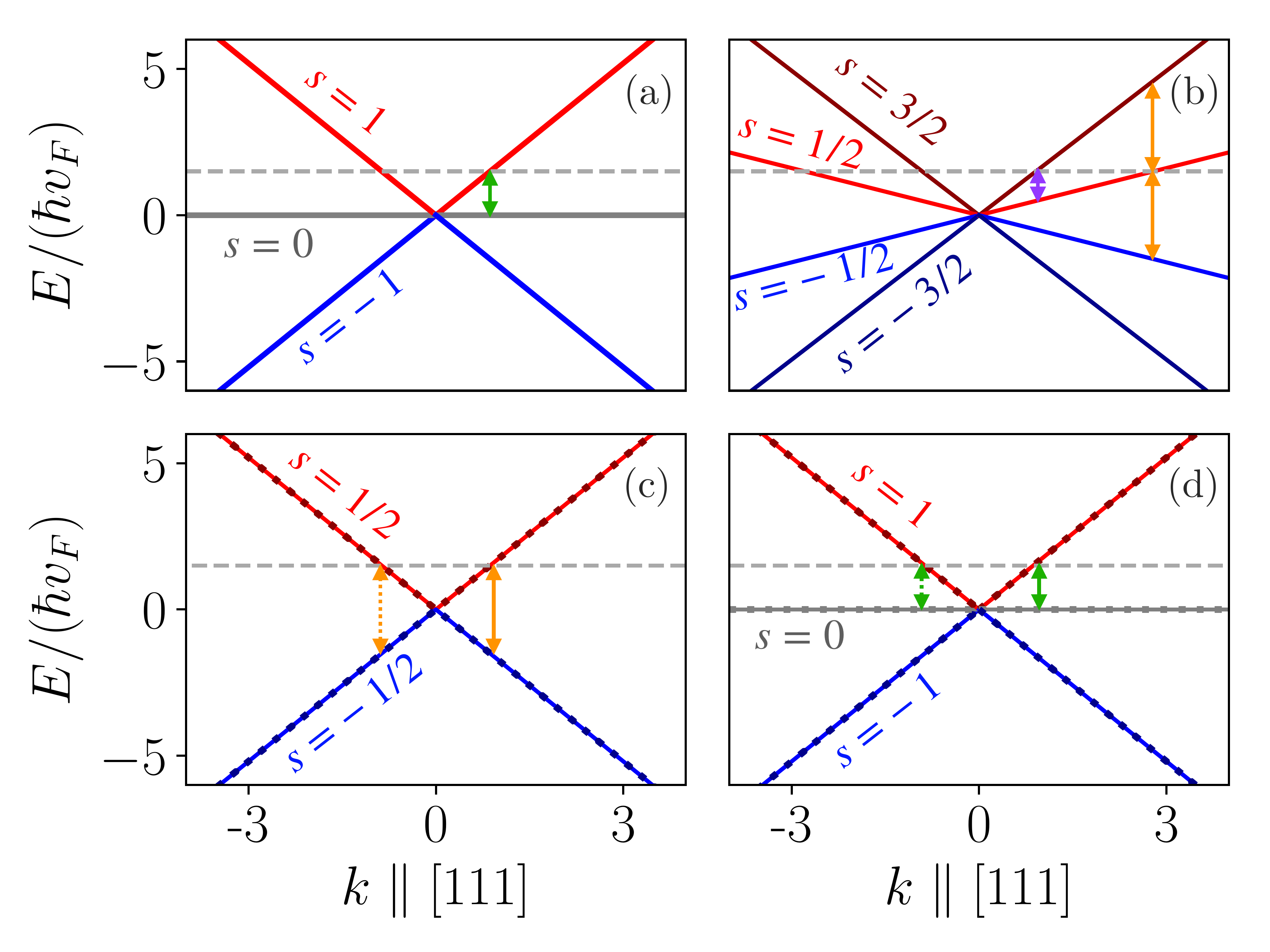}
    \caption{Band structures of the rotationally symmetric multifold fermions considered in Sec. \ref{subsec:lowenergysym} in the high-symmetry direction $\mathbf{k}^{111}=k(1,1,1)/\sqrt{3}$. (a) Threefold fermion (spin-$1$), (b) spin-$3/2$ fourfold fermion, (c) double spin-$1/2$ fourfold fermion and (d) sixfold fermion (double spin-$1$). The labels indicate the effective spin quantum number $s$ of each band. The vertical arrows indicate the only allowed interband transitions, those that satisfy $\Delta s = s-s'= \pm 1$, with characteristic frequencies $\hbar\omega=2\mu/3$ (purple), $\hbar\omega=\mu$ (green) and $\hbar\omega=2\mu$ (orange). The dotted lines in the double spin-$1/2$ and double spin-1 fermions indicate two degenerate copies of the spin-$1/2$ and spin-$1$ fermions, respectively, with the corresponding transitions indicated by dotted vertical arrows.}
    \label{fig:symmetricbands}
\end{figure}

The double spin-$1/2$ fourfold fermion consists of two decoupled copies of the Weyl Hamiltonian, and thus its optical conductivity is twice the optical conductivity of the Weyl fermion given by Eq.~\eqref{eqn:condweyl}, similar to Ref.~\onlinecite{roy_birefringent_2018-1}. We express it as $\sigma_{4f}^{2\times 1/2}(\omega)=2\sigma_{W}(\omega)$ (see Table~\ref{table:optconds} and Fig.~\ref{fig:symmetricbands}~(c)). If the Weyl bands are tilted, the characteristic frequency $\hbar\omega_W=2\mu$ at which the optical conductivity changes from being zero to being linear in $\omega$ depends on the magnitude of the tilt, but its linear dependence remains unaltered~\cite{mukherjee_doping_2018}.\\

We continue by considering the most general low energy Hamiltonian for a threefold fermion
\begin{equation}
\label{eqn:hamilt_3f}
H_{3f}(\mathbf{k},\phi)
=\hbar v_{F}
\begin{pmatrix}
0 & \me^{\mi\phi}k_x & \me^{-\mi\phi}k_y\\
\me^{-\mi\phi}k_x & 0 & \me^{\mi\phi}k_z\\
\me^{\mi\phi}k_y & \me^{-\mi\phi}k_z & 0
\end{pmatrix},
\end{equation}
where $v_F$ is the Fermi velocity and $\phi$ is a material-dependent parameter~\cite{bradlyn_beyond_2016,flicker_chiral_2018}. In the absence of spin-orbit coupling the value of $\phi$ is constrained to be $\phi_0=\pi/2 \mod(\pi/3)$~\cite{manes_existence_2012}. In this case the Hamiltonian takes the form $H_{3f}^{\phi_0}(\mathbf{k})\equiv H_{3f}(	\mathbf{k},\phi_0)= \hbar v_F \mathbf{k}\cdot \mathbf{S}_1$, where $\mathbf{S}_1$ is a vector of three spin-1 matrices which form a representation of SU(2) (see Appendix~\ref{app:symmetric}). The threefold fermions described by $H_{3f}^{\phi_0}$ have full rotational invariance and effective spin $S=1$, and we refer to them as symmetric threefold fermions.

The band energies for the spin-1 symmetric threefold fermion are $E_{s}= s \hbar v_F \vmod{\mathbf{k}}$ (see Fig.~\ref{fig:symmetricbands}~(a)), where $s=-1,0,1$ corresponds to the three possible values of the effective spin of the fermion. Because of this effective quantum number, a photon can excite an electron from a filled band $s$ to an unoccupied band $s'$ only if the selection rule $\Delta s= s'-s= \pm 1$ is satisfied, as depicted in Fig.~\ref{fig:symmetricbands}~(a). 

By inserting the analytic energies and the eigenfunctions of $H_{3f}^{\phi_0}$ (see Appendix~\ref{app:symmetric},  Eq.~\eqref{eqn:eigfuncs3f}) in Eq.~(\ref{eqn:optcond}) we obtain the optical conductivity
\begin{eqnarray}
\sigma_{3f}^{\phi_0}(\omega,\mu,\beta)=\frac{\omega e^2}{6\pi\hbar v_F}
\frac{\sinh(\hbar\omega\beta)}{\cosh(\hbar\omega\beta)+\cosh(\mu\beta)},
\label{eqn:optcond3ftemperature}
\end{eqnarray}
where the super-index $\phi_0$ refers to the symmetric case. 

Taking the $T\to 0$ ($\beta\to\infty$) limit, the optical conductivity simplifies to 
\begin{eqnarray}
\sigma_{3f}^{\phi_0}(\omega,\mu,\beta)=\frac{\omega e^2}{6\pi\hbar v_F}
\Theta(\hbar\omega-\mu).
\label{eqn:optcond3f}
\end{eqnarray}
From  Eq.~(\ref{eqn:optcond3f}), the optical conductivity of the threefold fermion is linear with $\omega$ as for the Weyl fermion, yet four times larger given the same Fermi velocity $v_F$ (see Table \ref{table:optconds}). Also, the characteristic frequency at which the optical conductivity starts to grow linearly with the frequency is $\hbar\omega_{3f}=\mu$, which is different from the characteristic frequency of the Weyl fermion $\hbar\omega_W=2\mu$. At $\omega = \omega_{3f}$ the only allowed interband transition is activated (green arrow in Fig.~\ref{fig:symmetricbands}~(a)), connecting a filled and an empty band with $\Delta s= s'-s= \pm 1$.

Since the low-energy Hamiltonian describing the sixfold fermion can be brought to a block-diagonal form with two copies of the threefold Hamiltonian in the diagonal, its optical conductivity is twice that of the threefold fermion (see Table~\ref{table:optconds} and Fig.~\ref{fig:symmetricbands}~(d)).\\

We now carry out a similar analysis to obtain the optical conductivity for the symmetric fourfold fermion. A fourfold degeneracy is found only with spin-orbit coupling in tetrahedral~\cite{chang_unconventional_2017,tang_multiple_2017} or octahedral~\cite{bradlyn_beyond_2016} subgroups~\cite{flicker_chiral_2018}.
A general fourfold fermion in the octahedral group has the Hamiltonian
\begin{equation}
\begin{aligned}
\label{eqn:ham_4fold}
H_{4f}&(\mathbf{k},a,b)=\\
&
\begin{pmatrix}
a k_z & 0 & -\frac{a+3b}{4}k_{+} & \frac{\sqrt{3}(a-b)}{4}k_{-}\\
0 & b k_z & \frac{\sqrt{3}(a-b)}{4}k_{-} &-\frac{3a+b}{4}k_{+}\\
-\frac{a+3b}{4}k_{-}& \frac{\sqrt{3}(a-b)}{4}k_{+} & -a k_z & 0\\
\frac{\sqrt{3}(a-b)}{4}k_{+} & -\frac{3a+b}{4}k_{-} & 0 & -b k_z
\end{pmatrix},
\end{aligned}
\end{equation}
where $k_\pm=k_x\pm ik_y$, and $a,b$ are two material-dependent parameters expressed in units of $\hbar v_F$, whose ratio we define as $\chi=\arctan(b/a)$. For tetrahedral groups,  an extra linear term is allowed, that we discuss in Appendix \ref{sec:4ftetra}.

A fourfold fermion recovers the full rotational symmetry when $\chi=\chi_0=\arctan(-3)$ ($b=-3a$) or $\chi=\chi_0=\arctan(-1/3)$ ($b=-a/3$), for which the Hamiltonian takes the form $H_{4f}^{\chi_0}(\mathbf{k})\equiv H_{4f}(\mathbf{k},\chi_0)=\hbar v_F \mathbf{k}\cdot \mathbf{S}_{3/2}$, where $\mathbf{S}_{3/2}$ are three matrices that form a spin-$3/2$ representation of SU(2) (see Appendix~\ref{app:symmetric}).

In this case, the energies are given by $E_s= 2s \hbar v_F \vmod{\mathbf{k}}$, with $s=-\tfrac{3}{2},-\tfrac{1}{2},\tfrac{1}{2},\tfrac{3}{2}$ corresponding to the effective spin of the multifold fermion (see Fig.~\ref{fig:symmetricbands}~(b)). Similar to the threefold case, the selection rules only allow transitions between a band $s$ and a band $s'$ such that $\Delta s= s'-s = \pm 1$.

\begin{figure}[t]
    \includegraphics[width=\linewidth]{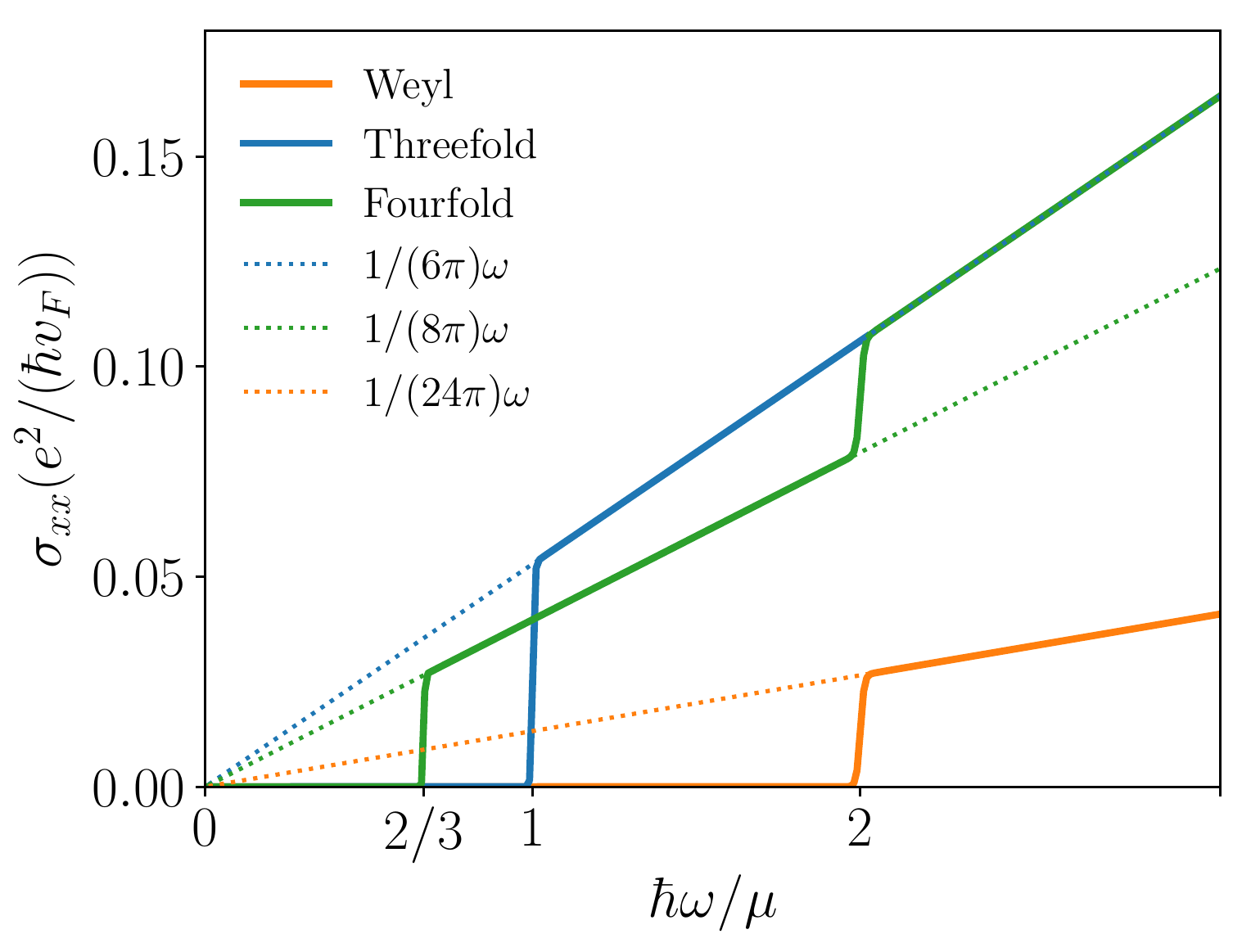}
    \caption{Comparison between the optical conductivities of the symmetric multifold fermions for which all cases in Table~\ref{table:optconds} are built. A single threefold or fourfold fermion has a larger conductivity than a Weyl fermion, normalized per node and by their Fermi velocity. Depending on the type of multifold the activation frequency can occur at $\hbar\omega=2\mu/3,\mu$ or $2\mu$.}
    \label{fig:Optcond_symmetric}
\end{figure}

\begin{figure*}
    \includegraphics[width=\textwidth]{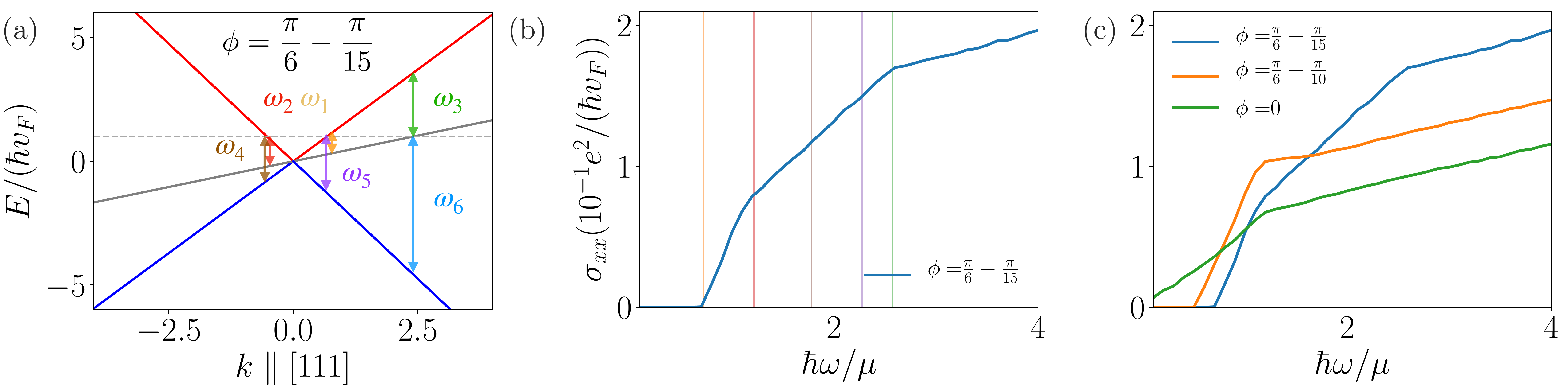}
    \caption{Non-symmetric threefold fermion. (a) Band structure for $\phi=\pi/6-\pi/15$ and the corresponding transitions allowed with their characteristic frequencies (the exact expressions for these are given in the Appendix~\ref{app:symmetric}, Eq.~\eqref{eqn:3f_frequencies}). (b) Optical conductivity for the non-symmetric threefold fermion depicted in (a). The characteristic frequencies are represented by vertical lines with colors corresponding to the transitions depicted in (a). The frequencies $\omega_3$ and $\omega_4$ do not affect the optical conductivity since they correspond to transitions with $\Delta s\neq \pm 1$, which are forbidden for the symmetric case. (c) Optical conductivities of non-symmetric threefold fermions for different values of the material dependent parameter $\phi$.}
    \label{fig:3fnosym}
\end{figure*}

\begin{figure*}
    \includegraphics[width=\textwidth]{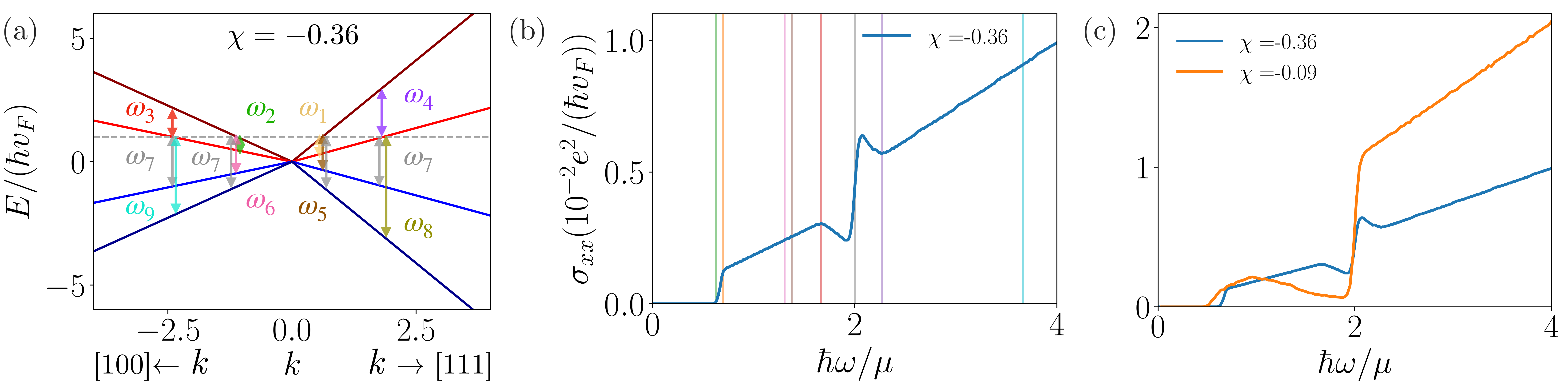}
    \caption{Non-symmetric fourfold fermion. (a) Band structure for $\chi=-0.36$ ($a=3.2$, $b=-1.2$) and the corresponding transitions allowed with their characteristic frequencies (the exact expressions for these are given in the Appendix~\ref{app:symmetric}, Eq.~\eqref{eqn:4f_frequencies}). (b) Optical conductivity for the non-symmetric fourfold fermion depicted in (a). The characteristic frequencies are represented by vertical lines with colors corresponding to the transitions depicted in (a). The frequencies that do not affect the optical conductivity correspond to transitions with $\Delta s\neq \pm1$, which are forbidden for the symmetric case. (c) Optical conductivities of non-symmetric fourfold fermions for different values of the material dependent parameter $\chi=-0.36$ ($a=3.2$, $b=-1.2$) and $\chi=-0.09$ ($a=3.4$, $b=-0.3$).}
    \label{fig:4fnosym}
\end{figure*}

Inserting the energies and the eigenfunctions, which can be obtained analytically, in the expression for the optical conductivity in Eq.~\eqref{eqn:optcond} we obtain 

\begin{eqnarray}
\label{eqn:cond4ftemperature}
\nonumber
\sigma_{4f}^{\chi_0}(\omega,\mu,\beta)&=&\frac{\omega e^2}{8\pi\hbar v_F}\left[\frac{\sinh(\hbar\omega\beta/2)}{\cosh(\hbar\omega\beta/2)+\cosh((\mu-\hbar\omega)\beta)}\right.\\
&+& \left.\frac{4}{3}\frac{\sinh(\hbar\omega\beta/2)}{\cosh(\hbar\omega\beta/2)+\cosh(\mu\beta)}\right].
\end{eqnarray}
Taking the zero temperature limit $T\to 0$ Eq.~\eqref{eqn:cond4ftemperature} is simplified considerably to 
\begin{eqnarray}
\label{eqn:cond_4f}
\sigma_{4f}^{\chi_0}(\omega,\mu)=\frac{\hbar\omega e^2}{8\pi\hbar v_F}\left[\frac{1}{3}\Theta(\hbar\omega-2\mu)+ \Theta(\hbar\omega-\frac{2}{3}\mu)\right].
\end{eqnarray}

As in the threefold case, the conductivity is linearly dependent on the frequency $\omega$ of the photon. In this case we find two characteristic frequencies due to the more complex band structure, $\hbar\omega_{4f,1}=2\mu/3$ and $\hbar\omega_{4f,2}=2\mu$ (see Fig.~\ref{fig:symmetricbands}~(b)), defining two separated regions in the optical conductivity with different linear dependence on $\omega$. 
When $\omega_{4f,1}<\omega$ one transition with $\Delta s =\pm 1$ from the intermediate-upper band to the upper band is allowed, until it vanishes at $\omega=\omega_{4f,2}$. When $\omega >\omega_{4f,2}$ a transition between the two intermediate bands is activated (lower orange arrow in Fig.~\ref{fig:symmetricbands}~(b)).

In Fig.~\ref{fig:Optcond_symmetric} we compare the optical conductivities of the twofold (Weyl) fermion and the symmetric threefold and fourfold fermions discussed in this section. For $2\mu/3<\hbar\omega<2\mu$ the optical conductivity of the spin-$3/2$ fourfold is larger than that of the Weyl for a given $v_F$, but smaller than that of the threefold, while in the region $\hbar\omega>2\mu$ the optical conductivity of the threefold and the fourfold are equal.
The characteristic frequencies that activate the interband transitions identify each symmetric multifold fermion, and they do not depend on dimensionality~\cite{Dora}. Similarly, the ratio between the symmetric multifold optical conductivities shown in Fig.~\ref{fig:Optcond_symmetric} is the same\footnote{The ratio shown in Ref.~\onlinecite{Dora} is recovered by choosing the same convention that they present for the spin operators. Here we do not include the spin factor for half-integer spins.} as for two-dimensional multifold systems~\cite{Dora}.

The abruptness of the jump in the optical conductivity at the characteristic frequencies depends on the temperature. Thermally activated carriers will populate states above the Fermi level and empty states below it, smoothing the step function in Eq.~\eqref{eqn:optcond3f} (see Appendix~\ref{sec:smoothing}, Fig.~\ref{fig:temperature}). Additionally, the presence of disorder introduces a finite scattering time $\tau$ resulting in a finite $\delta=1/\tau$ in Eq.~\eqref{eqn:optcond}. In the simplest approximation, where $\tau$ is a constant, the step function will be broadened\cite{ashby_magneto-optical_2013}, similar to the finite temperature case discussed in Appendix~\ref{sec:smoothing}.

\subsection{Optical conductivity in non-symmetric low energy models}

In real materials, $\phi$ and $\chi$ are pinned to the symmetric values $\phi_0$ and $\chi_0$ only if spin-orbit coupling is absent. Including spin-orbit coupling for a particular multifold splits it into multifolds at the same high-symmetry point but with different degeneracy. For  example, in space group 198 a threefold at $\Gamma$ splits into one fourfold fermion and one Weyl fermion. This is general: multifolds without spin-orbit coupling are spinless and have $\phi=\phi_0$ or $\chi=\chi_0$, while spinful multifolds may have any value of these parameters and occur in different high symmetry points compared to the spinless case.

In particular, for a generic threefold fermion occurring in the presence of spin orbit coupling the material-dependent parameter is no longer restricted to $\phi=\phi_0$, and can take values in the range $\pi/3<\phi<2\pi/3 \mod \pi/3$~\cite{bradlyn_beyond_2016}. A change in $\phi$ will tilt the bands, breaking the full rotational symmetry. In this case, the selection rules of the symmetric model no longer apply and more excitations are allowed, as depicted in Fig.~\ref{fig:3fnosym}~(a), since the effective spin is no longer a good quantum number.
The characteristic frequencies $\omega_i(\phi)$ associated to each transition depicted in Fig.~\ref{fig:3fnosym}~(a) can be obtained analytically~\cite{flicker_chiral_2018} and we reproduce them for completeness in Appendix~\ref{app:symmetric}.

The activation of new transitions at each $\omega_i$ results in a change in the linear dependence on $\omega$ of the optical conductivity, as depicted in Fig.~\ref{fig:3fnosym}~(b).
Some transitions have a large effect on the slope, while others barely affect it. This is consistent with other optical effects in multifold fermions~\cite{flicker_chiral_2018} and is rooted in the fact that the matrix elements for transitions with $\Delta s \neq \pm 1$ are typically smaller than those with $\Delta s = 1$. 
In Fig.~\ref{fig:3fnosym}~(c) we plot the optical conductivity for different values of $\phi$. Changing this parameter shifts the characteristic frequencies according to their analytic expression $\omega_i(\phi)$, given in Eq.~\ref{eqn:3f_frequencies}. As apparent in Fig.~\ref{fig:3fnosym}~(c), the slope of the optical conductivity also depends on $\phi$, yet we find no closed analytic form. 

Combining all the results, we find that it is possible to identify a generic threefold fermion in an optical experiment, provided $\phi$ and $v_F$ are known (for example either from first principles calculations or photemission data).

We find a similar behavior in the fourfold case. For an arbitrary value of $\chi\neq\chi_0$ we lose full rotational symmetry and the spin-$3/2$ picture breaks down, allowing for new electronic excitations in the system (see Fig.~\ref{fig:4fnosym}~(a)). The characteristic frequencies for these excitations can be obtained analytically~\cite{flicker_chiral_2018} (see Eq.~\eqref{eqn:4f_frequencies}), and produce a change in the linear dependence on $\omega$ of the optical conductivity, as we see in Fig.~\ref{fig:4fnosym}~(b) and (c).\\

The characteristic frequencies at which the optical conductivity changes and the linear dependence on $\omega$ are different for each multifold, which allows us to identify them by their optical conductivity for both symmetric and non-symmetric cases. 

\begin{figure*}
    \includegraphics[width=\linewidth]{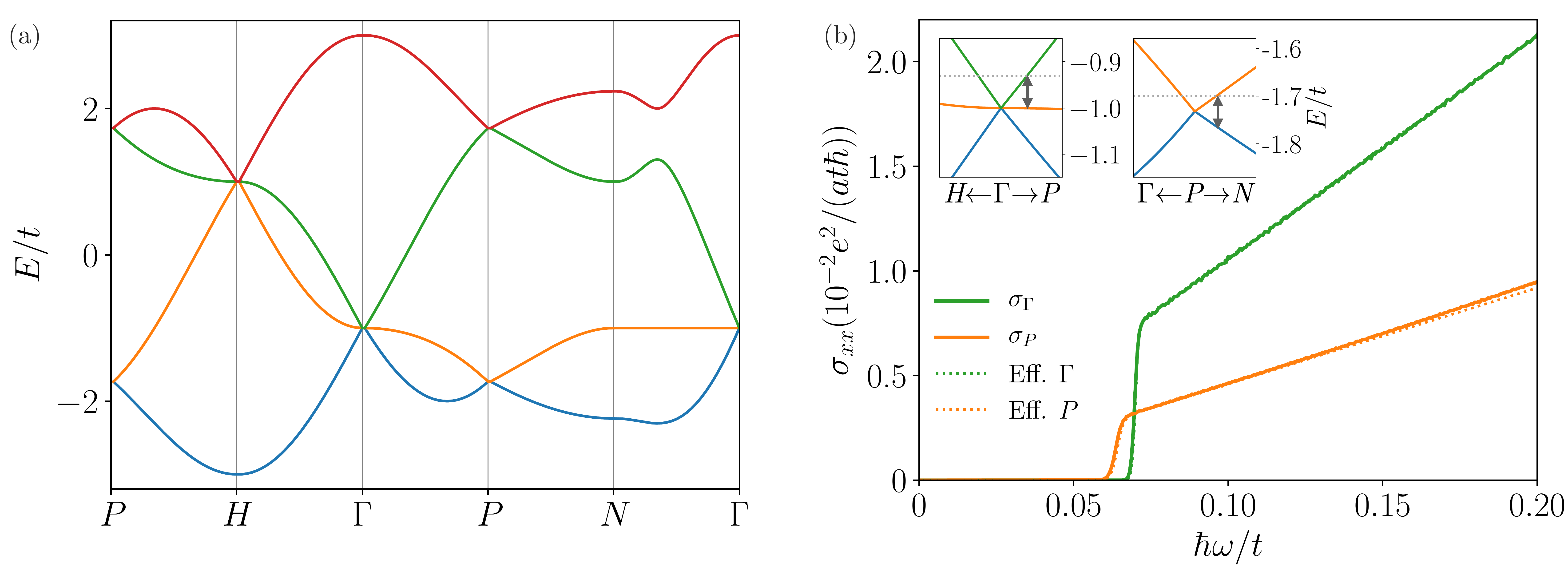}
    \caption{Tight-binding model for a material in SG199. (a) Band structure of the tight-binding model used in Sec. \ref{subsec:SG199} obtained from Refs.~\onlinecite{chang_unconventional_2017,flicker_chiral_2018}. (b) Optical conductivity of the tight-binding model calculated with a chemical potential $\mu/t=-0.93$ (solid green line), separating by $0.07$ the threefold node at the $\Gamma$ point and the Fermi level (left inset). In solid orange the optical conductivity calculated for the tight-binding model with $\mu/t=-1.7$, separating by $0.032$ the lower Weyl node at the $P$ point and the Fermi level (right inset). We present the optical conductivity of the effective models described in \ref{subsec:SG199} for the $\Gamma$ point (dashed green) obtained with Eq.~\eqref{eqn:optcond3ftemperature}, and for the $P$ point (dashed orange) obtained with Eq.~\eqref{eqn:condweyl}. In the frequency range $0<\hbar\omega/t<0.2$ the optical conductivity is well described by the linear effective model, i.e. the solid and dashed lines fall on top of each other. The orbital embedding does not affect the results at these energy scales. These results are obtained with $1/\beta=5\times 10^{-4}t$.}
    \label{fig:panel_199}
\end{figure*}

\subsection{Imaginary part of the optical conductivity and sum rules}

Before discussing realistic tight-binding models we note that so far we have calculated only the absorptive (real) part of the optical conductivity. Using the Kramers-Kronig transformations~\cite{dresselhaus_optical_nodate} we have obtained the dispersive (imaginary) of the optical conductivity in Appendix~\ref{sec:KK}, where we derive a general expression applicable to all symmetric and non-symmetric cases, and we compute it explicitly for the symmetric cases. 

For completeness, in Appendix~\ref{sec:sumrules} we compute the conductivity sum rule. The sum rule relates the integral over all frequencies of the real part of the optical conductivity, $\braket{\sigma}$, to the total number of particles. Since low energy linearly dispersing bands, such as those of Weyl or multifold fermions, are unbounded, the f-sum rule explicitly depends on the cut-off scale $\Lambda$, similar to what is known for graphene~\cite{sabio_f-sum_2008,Ando}. Leaving the closed form and details to Appendix~\ref{sec:sumrules}, we simply mention that for symmetric multifolds the sum rule of the interband part of the conductivity takes the form $\braket{\sigma}\propto (\Lambda^2-c\mu^2)$ where $c$ is a factor that depends on the type of multifold. Specifically $c=1$ and $c=4/3$ for the symmetric threefold and fourfold cases respectively.

\section{Optical conductivity of Multifold Fermions: realistic models}
\label{sec:realistic}

The fingerprints of chiral multifold fermions in the optical conductivity allow us to identify them also in real materials. To make material-specific predictions 
we use tight-binding models with parameters that reproduce first principle band structures of space groups SG199 and SG198~\cite{flicker_chiral_2018,chang_unconventional_2017,Pshenay_Severin_2018}, that realize all types of chiral multifold fermions. 

The tight-binding models that we use capture specific properties of the material, such as the energy scales, the band connectivity and multifold crossings, and the orbital embedding. The latter describes the spatial position (or embedding) of the orbitals in real space. A change in the orbital embedding acts as a momentum dependent unitary transformation of the tight-binding Hamiltonian: it does not modify the band structure of the material, but modifies its eigenfunctions. It is thus necessary to take it into account to give accurate predictions of observables, in particular the optical conducitivity. The details of this transformation depend on the space group, and we present the explicit form of the Hamiltonians with orbital embedding for SG199 and SG198 in Appendix \ref{sec:orbitalembedding}. 

\subsection{Space Group 199}
\label{subsec:SG199}
The first realistic tight-binding model that we consider describes a material in SG199 without spin-orbit coupling, which captures the adequate band connectivity and chirality. Since no material has been found in this space group with only multifold fermions near the Fermi level~\cite{bradlyn_beyond_2016} we present the results for this model in units of the characteristic hopping scale $t>0$ and the lattice constant $a$. If we parametrize the orbital embedding by a scalar $u$, a generic value in the range $-1/2<u<1/2$ sets the model to be in SG199. Choosing $u=1/4$ increases the symmetry from tetrahedral to octahedral, provided the hoppings do not break this symmetry, describing a material in SG214. These requirements are satisfied by our tight-binding model and thus it can interpolate between SG199 and 214 depending on the value of $u$. The explicit expression for the tight-binding model and its embedding can be found in Appendix~\ref{sec:orbitalembedding}.

In Fig.~\ref{fig:panel_199} (a) we show a representative band structure of a material in SG199. It features protected threefold nodes at the $\Gamma$ point at energy $\mu_{\Gamma}/t=-1$ and at the $H=(-\pi,\pi,\pi)$ point with $\mu_{H}/t=1$. It also hosts two Weyl nodes at the $P=(\pi/2,\pi/2,\pi/2)$ point, at energies $\mu_{W_1}/t=-1.732$ and $\mu_{W_2}/t=1.732$. 

To focus on the optical conductivity of the threefold fermion in SG199, we can place the chemical potential slightly above the threefold node at the $\Gamma$ point, at $\mu/t=-0.93$. We present the conductivity for this case in Fig.~\ref{fig:panel_199} (b). It has a linear dependence on the frequency $\omega$ and exhibits a change in the slope at $\hbar\omega_{\Gamma}/t=0.07$. 
This result matches exactly the analytic results obtained for a threefold fermion in Eq.~\eqref{eqn:optcond3ftemperature} in two ways. First the activation frequency $\hbar\omega_{3f}/t=\mu/t=0.07$ exactly matches the distance from the node to the Fermi surface. Second, the numerical slope coincides with the slope determined by the effective Fermi velocity that we obtain by projecting the tight-binding Hamiltonian on the three eigenstates corresponding to the $\Gamma$ point. This projection can be brought to the form of the threefold model in Eq.~\eqref{eqn:hamilt_3f} with a unitary transformation~\cite{manes_existence_2012}, with an effective Fermi velocity $v_F=a t/(2\hbar)$, where $a$ is the lattice constant and $t$ is the hopping parameter in the tight-binding model.

If we instead place the chemical potential at $\mu/t=-1.7$, near the lower Weyl node at energy $\mu_{W2}$ around $P$, we can focus on the optical conductivity of this Weyl node. We can see in Fig.~\ref{fig:panel_199}~~(c) that it has a linear dependence on the frequency $\omega$ and a change in the slope at $\hbar\omega_P/t=0.064$. This energy scale matches that of a Weyl fermion (see Table~\ref{table:optconds}) with an activation frequency of  $\hbar\omega_{W}/t=2\mu/t=0.064$, corresponding to twice the distance from the node to the Fermi surface. The slope matches that of Eq.~\eqref{eqn:condweyl} using the effective Hamiltonian around the $P$ point. We obtain this model by projecting the Hamiltonian on the corresponding eigenstates near the Weyl node and bringing it to a Weyl Hamiltonian form $H=\hbar v_F\mathbf{k}\cdot\boldsymbol{\sigma}$ with a unitary transformation,  where $v_F=at/(2\sqrt{3}\hbar)$\cite{manes_existence_2012}.

\subsection{Space Group 198: RhSi}
\label{subsec:SG198}

\begin{figure*}
    \includegraphics[width=\linewidth]{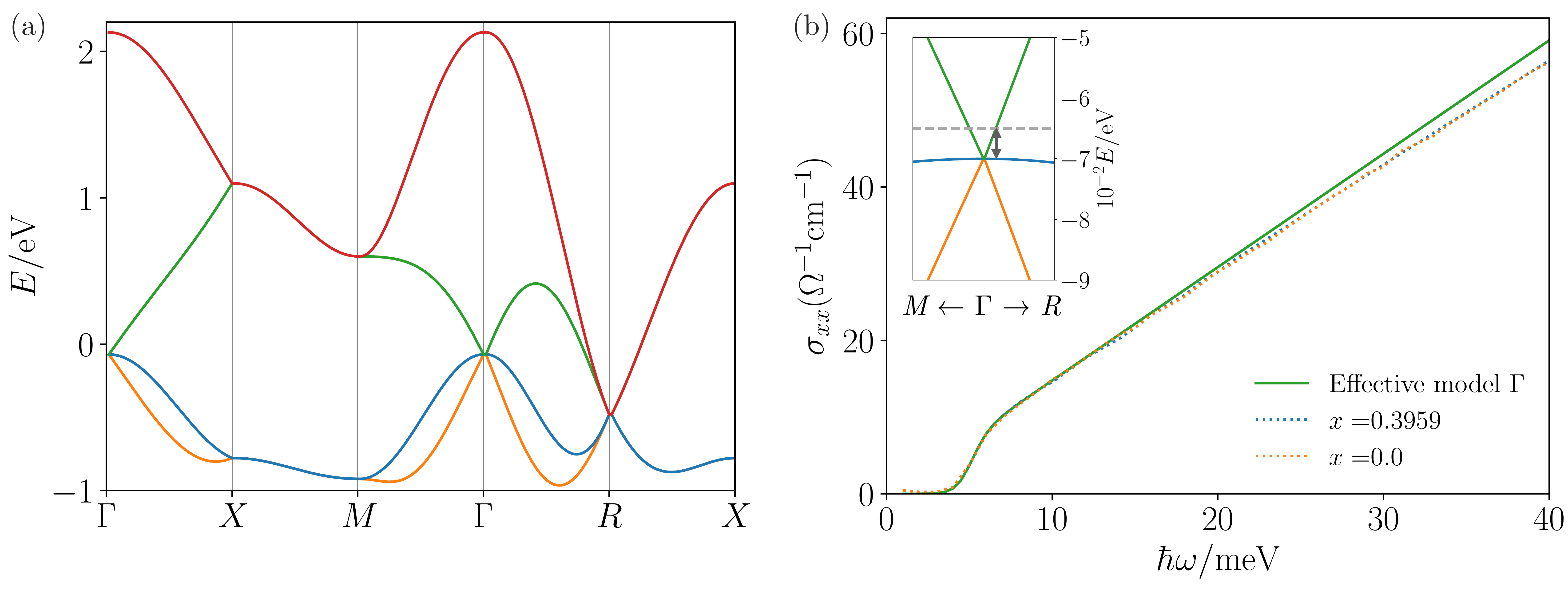}
    \caption{(a) Band structure of the tight-binding model of RhSi used in Sec.~\ref{subsec:SG198} obtained from Refs.~\onlinecite{chang_unconventional_2017,flicker_chiral_2018}. (b) Optical conductivity corresponding to excitations near the $\Gamma$ point calculated for $1/\beta=0.5~\mathrm{meV}$ ($T=5.8~\mathrm{K}$), including the spin degeneracy and $\mu=65~\mathrm{meV}$. The latter sets an energy difference of $5~\mathrm{meV}$ between the threefold node at the $\Gamma$ point and the Fermi level (inset). The results without taking into account the orbital embedding (dashed orange line) and with the orbital embedding for RhSi (dashed blue line) are close in the range of frequencies plotted, $0<\omega<40~\mathrm{meV}$. The numerical results obtained for the tight-binding model (dashed lines) are similar to the optical conductivity of the effective model at $\Gamma$ (green line) discussed in Sec.~\ref{subsec:SG198} for $\omega\lesssim 12~\mathrm{meV}$, and exhibit a jump at $\omega=5~\mathrm{meV}$, a characteristic of the threefold fermion. For higher frequencies the linear effective model fails to capture the curvature of the bands where higher-order terms become important, causing the optical conductivity to deviate from that of the tight-binding model.}
    \label{fig:zoom198}
\end{figure*}

\begin{figure}
    \includegraphics[width=1\linewidth]{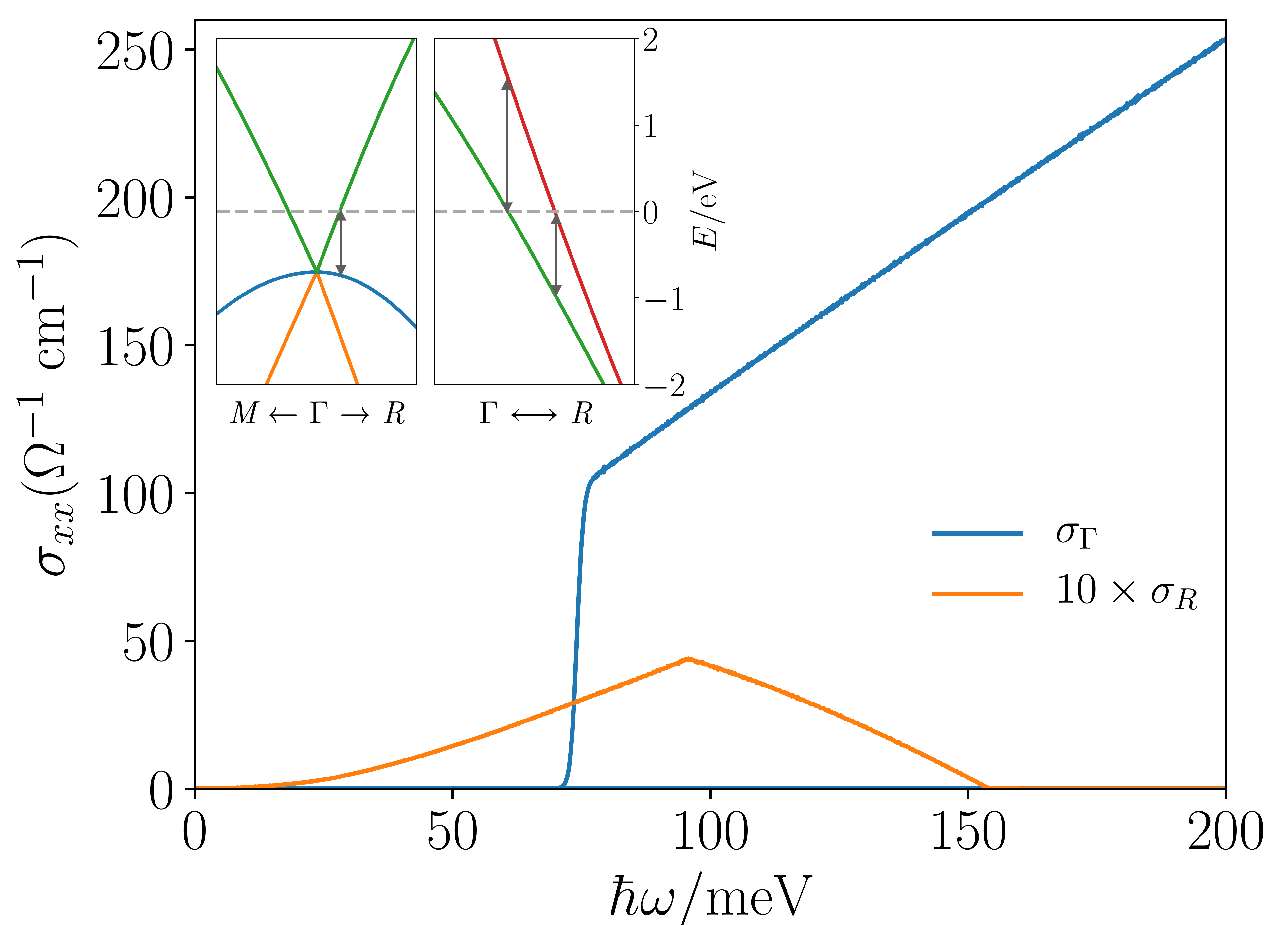}
    \caption{Optical conductivity of RhSi obtained using the tight-binding model of SG198 described in Refs.~\onlinecite{chang_unconventional_2017,flicker_chiral_2018} including the spin degeneracy, with $\mu=0~\mathrm{eV}$ and $1/\beta=0.5~\mathrm{meV}$ ($T=5.8~\mathrm{K}$). The contribution of the $\Gamma$ point (blue line) is activated by excitations between the intermediate band (left inset, blue line) to the upper band (left inset, green line), and exhibits a jump near $\omega=74~\mathrm{meV}$, which is larger than $|\mu_{3f}|=70~\mathrm{meV}$, set by the concavity of the intermediate band. The contribution due to excitations near the $R$ point (orange line) is magnified by a factor 10 for comparison. This contribution is activated by transitions between the intermediate-upper (green) and upper band (red), depicted in the right inset. The characteristic frequencies, represented by vertical arrows in the right inset, correspond to the maximum value of $\sigma_R$ at $\omega=96$ meV and its vanishing at frequency $\hbar\omega=154$ meV.}
    \label{fig:RhSi}
\end{figure}

The next model that we consider describes a material in SG198. A variety of materials in this space group have been theoretically predicted to be chiral multifold semimetals~\cite{chang_unconventional_2017,tang_multiple_2017,bradlyn_beyond_2016,Pshenay_Severin_2018} and these expectations have been confirmed by angle resolved photoemission in RhSi~\cite{sanchez_discovery_2018}, CoSi~\cite{takane_observation_2018,Rao2018} and AlPt~\cite{schroter_topological_nodate}. In this section we calculate the optical conductivity of RhSi as a representative material in SG198. In order to do so, we use the model originally presented in Ref.~\onlinecite{chang_unconventional_2017} for RhSi, whose hopping parameters are fitted to first principle band calculations. We upgrade this model as in Ref.~\onlinecite{flicker_chiral_2018}: we take into account the orbital embedding by conjugating the tight-binding Hamiltonian with a unitary matrix parametrized by $x$, with $x=0.3959$ for RhSi. Further details of this model can be found in Appendix~\ref{sec:orbitalembedding}.

In Fig.~\ref{fig:zoom198}~(a) we present the band structure of RhSi without spin-orbit coupling, where we chose the zero of energies to coincide with the predicted Fermi level of RhSi. It exhibits a protected threefold crossing at the $\Gamma$ point at $\mu_{3f}=-0.07~\mathrm{eV}$ and a protected fourfold crossing (double spin-$1/2$) at the $R=(\pi,\pi,\pi)$ point at $\mu_{4f}=-0.48~\mathrm{eV}$. 

Before studying the realistic optical conductivity of RhSi it is instructive to place the chemical potential close to the threefold at $\Gamma$ ($\mu=0.065~\mathrm{eV}$) to compare it with the optical conductivity of the linear low energy model. In Fig.~\ref{fig:zoom198} (b) we present the results obtained numerically choosing the orbital embedding for RhSi ($x=0.3959$), the results without orbital embedding ($x=0$) and the analytic results for the effective model obtained following the projection procedure described for SG199 in the previous section.
As for SG199 the projection around $\Gamma$ results in the effective Hamiltonian Eq.~\eqref{eqn:hamilt_3f} with
$v_F = a t/(2\hbar)$, where $a=4.6$~\AA~ for RhSi and with $t=0.76$ eV chosen to match the multifold low energy bands~\cite{chang_unconventional_2017}. Fig.~\ref{fig:zoom198} (b) shows that the numerical results match the optical conductivity of the effective model for $\omega \lesssim 12~\mathrm{meV}$, they grow linearly with $\omega$ and have a step at $\omega_{\Gamma}=5~\mathrm{meV}$, which is the energy separation from the node to the Fermi surface.  
For $\omega\gtrsim 12~\mathrm{meV}$ the quadratic corrections become important, and the optical conductivity calculated with the tight-binding model departs from the linear dependence obtained for the effective model. At the same scale, the results obtained for $x=0$ and $x=0.3959$ do not match exactly, which indicates that the higher-order corrections are sensitive to the orbital embedding unlike the linear approximation.

We now consider the actual values of the chemical potential and the orbital embedding that describe RhSi, which are $\mu=0$ and $x=0.3959$ respectively. We recall that $\mu=0$, as set by \textit{ab-initio} calcualtions~\cite{chang_unconventional_2017}, lies $0.07~\mathrm{eV}$ above the threefold fermion at $\Gamma$, and $0.48~\mathrm{eV}$ above the fourfold node at the $R$ point. 

For these material parameters, and in the $0<\hbar\omega< 200~\mathrm{meV}$ frequency range, the interband optical conductivity has contributions from transitions close to the $\Gamma$ and $R$ points, that we present separately in Fig.~\ref{fig:RhSi}. The contribution to the conductivity near the $\Gamma$ point exhibits a jump at a frequency $\hbar\omega_{\Gamma}=\mu_{3f}=74~\mathrm{meV}$, which is slightly larger than the corresponding characteristic frequency of a threefold fermion $\hbar\omega_{3f}=70~\mathrm{meV}$ (see Table~\ref{table:optconds}). This is due to the curvature of the intermediate band, which results in a higher activation frequency for the allowed transition near the $\Gamma$ point (see left inset in Fig.~\ref{fig:RhSi}).

Near the $R$ point, the only transitions that contribute below $\hbar\omega < 4\mu_f\sim 1~\mathrm{eV}$ are the interband transitions from the intermediate-upper band (green) to the upper band (red), that we depict in Fig.~\ref{fig:RhSi}, right inset. Their contribution to the conductivity is two orders of magnitude smaller compared to that associated to the $\Gamma$ point (see Fig.~\ref{fig:RhSi}). This small magnitude is to be expected once we recall that at low energies, near the node at $R$, these two bands correspond to two decoupled Weyl fermions  (see Fig.~\ref{fig:zoom198}~(a)), and the transitions between them are forbidden. As we increase the energy, the matrix elements grow as the bands separate. Since the separation is small, the matrix elements are small. The two extremal energies, depicted by the arrows in the right inset of Fig.~\ref{fig:RhSi}, correspond to the frequencies $\hbar\omega=96$ meV and $\hbar\omega=154$ meV, which match the scales where the $R$ point conductivity reaches its maximum and vanishes respectively (see Fig.~\ref{fig:RhSi}).

In summary, the interband optical conductivity of RhSi in the frequency range $\hbar\omega < 200$ meV is determined by that of the threefold fermion at the $\Gamma$ point, since the contribution of the fourfold at the $R$ point is two orders of magnitude lower.

\section{\label{sec:conclusions}Conclusions}

In this work we have shown that, per node, multifold semimetals have larger optical conductivity than Weyl semimetals. They also feature characteristic activation frequencies that are specific to each class of multifold degeneracy. These activation frequencies, as well as the slope of the conductivity as a function of frequency, can be used as a fingerprint to distinguish each chiral multifold crossing. We have considered multifold fermions in rotationally symmetric and non symmetric cases and realistic hamiltonians in space groups 199 and 198. RhSi, CoSi and AlPt~\cite{sanchez_discovery_2018,Rao2018,takane_observation_2018,schroter_topological_nodate} belong to the latter space group and thus our predictions can be readily tested in experiment. Our results complement known results for other topological semimetallic systems~\cite{carbotte_optical_2006,ahn_optical_2017,mukherjee_doping_2018,roy_birefringent_2018-1}. 

Partially motivated by recent optical experiments~\cite{Rees:2019ue} we have focused our material discussion on RhSi. 
In this material, without spin-orbit coupling, the interband optical conductivity is dominated by the electronic excitations of the threefold band crossing at the $\Gamma$ point, activated for frequencies above $74$ meV. The interband contribution of the $R$ point is negligible compared to that of the $\Gamma$ point. 

In experiments, the intraband Fermi surface contribution can mask some characteristics of the contribution of $\Gamma$ at low frequencies. In the presence of weak disorder the Drude peak is broadened by a scale set by the inverse scattering time $1/\tau$, estimated to be $\tau\sim$ ps ($\hbar/\tau \sim 10$ meV) for typical topological semimetals. Nevertheless, the Drude-like intraband contribution can be fitted with a Lorentzian distribution and subtracted in the experimental data analysis, revealing the characteristic features of the multifold fermions. Additionally, the tight-binding model we have used can underestimate the importance of the trivial pocket at $M$ at the Fermi level for some materials in SG198, such as AlPt but most likely not RhSi. Therefore we expect that for sufficiently clean samples of RhSi at low temperatures the Drude peak can be narrow enough to observe all the features described in this work. 

When considering realistic tight-binding models, we have not included spin-orbit coupling. In SG 198, for example, spin-orbit coupling splits the threefold fermion at the $\Gamma$ point into a fourfold (spin-3/2) fermion and a Weyl fermion. The fourfold at $R$ splits into a sixfold fermion and a Weyl fermion. The splitting scale is determined by the spin-orbit coupling energy scale. However, this splitting is too small ($\sim$ meV) to be observed in ARPES measurements in CoSi, RhSi and AlPt~\cite{sanchez_discovery_2018,takane_observation_2018,Rao2018,schroter_topological_nodate} and in recent optical conductivity data in RhSi~\cite{Rees:2019ue}. These observations justify our approximation and motivate future optical experiments with meV resolution.

From Fig.~\ref{fig:RhSi} we predict that RhSi has an optical conductivity at $\hbar\omega=0.1$ eV of $\sigma \sim 120$ $\Omega^{-1}\mathrm{cm}^{-1}$ determined by the threefold fermion at $\Gamma$. Unfortunately, a dedicated optical conductivity experiment for any of the above multifold materials is still lacking. However,  Ref.~\onlinecite{Rees:2019ue} recently reported that in the range $0.5$ eV $ \lesssim \hbar\omega\lesssim 0.8$ eV the conductivity of RhSi falls in the interval  $350~\Omega^{-1}\mathrm{cm}^{-1}\lesssim\sigma \lesssim~500~\Omega^{-1}\mathrm{cm}^{-1}$. To compare with these measurements we have calculated the optical conductivity in this range of frequencies and at $\hbar\omega=0.5~\mathrm{eV}$ we find $\sigma \sim 650$ $\Omega^{-1}\mathrm{cm}^{-1}$. At such high energies, there are several factors that can lead to this discrepancy. These include inaccuracies of the estimated value of the embedding $x$ or the tight-binding hopping parameters, as well as active transitions in other pockets such as those at $M$. At low energies, tight-binding models become more accurate and the effect of the orbital embedding is less relevant. Therefore, we expect that experiments carried out at lower frequencies would agree better with the expectations of our calculations.   

Our predictions are of special relevance to interpret the recent optical measurements of non-linear circular photocurrents in RhSi~\cite{Rees:2019ue}, and in particular to determine the topological monopole node charge from this measurement. This is because in practice, a good knowledge of the linear optical conductivity is important to interpret non-linear optical experiments~\cite{Patankar2018,Rees:2019ue}.
First, the absorption of the material determines the total non-linear current that can be measured through the glass coefficient, which is the ratio between non-linear current density and the absorption. Second, dissipative non-linear effects depend on the optical scattering time $\tau$. The linear optical conductivity can be used to estimate the magnitude of $\tau$, for example by quantifying a finite conductivity in the Pauli blocked region~\cite{Grushin2009}.  This estimate can then be used to assess the accuracy of the expected quantization of injection currents in mirror free semimetals~\cite{de_juan_quantized_2017,flicker_chiral_2018,Rees:2019ue}.

Our results show that the optical conductivity distinguishes the type of chiral multifold fermions in real materials and that it can be larger, per node, than a single Weyl fermion. We expect that our analysis of realistic models helps to interpret upcoming optical experiments in different multifold candidate materials, especially those in SG198, such as RhSi, CoSi and AlPt. 

\section{Acknowledgments}
The authors are indebted to B. Bradlyn, F. Flicker, S. Fratini, T. Morimoto and M. Vergniory for related collaborations and valuable comments. We thank M. Orlita for critical reading of the manuscript. We acknowledge support from the European Union's Horizon 2020 research and innovation programme under the Marie-Sklodowska-Curie grant agreement No. 754303 (M. A. S. M.) and 653846 (A. G. G) and the GreQuE Cofund programme (M. A. S. M). A. G. G. is also supported by the ANR under the grant ANR-18-CE30-0001-01.

\appendix

\section{Eigenfunctions and characteristic frequencies for the symmetric threefold and fourfold fermions\label{app:symmetric}}

\subsection{Threefold fermion}

For $\phi=\phi_0=\pi/2$ we can write the Hamiltonian for a threefold fermion (see Eq.~\eqref{eqn:hamilt_3f}) as $H_{3f}^{\phi_0}(\mathbf{k})=\hbar v_F \mathbf{k}\cdot \mathbf{S_1}$, where $\mathbf{S_1}=(S_{1,x},S_{1,y},S_{1,z})$ are the spin-1 matrices 
\begin{align}
\label{eqn:spin1matrices}
S_{1,x}=&
\begin{pmatrix}
0 & \mi & 0\\
-\mi& 0 & 0\\
0 & 0 & 0
\end{pmatrix},\nonumber \\
S_{1,y}=&
\begin{pmatrix}
0 & 0 & -\mi\\
0 & 0 & 0\\
\mi & 0 & 0
\end{pmatrix}, \nonumber \\
S_{1,z}=& 
\begin{pmatrix}
0 & 0 & 0\\
0 & 0 & \mi\\
0 & -\mi & 0
\end{pmatrix},
\end{align}
with commutation relations $\left[S_{1,i},S_{1,j}\right]=-i\epsilon_{ijk}S_{1,k}$.

The eigenstates of the threefold low energy model in Eq.~\eqref{eqn:hamilt_3f} were previously obtained analytically for any value of $\phi$ (see for instance Refs.~\onlinecite{bradlyn_beyond_2016,flicker_chiral_2018}),
\begin{align}
\label{eqn:eigfuncs3f}
\psi_s = \frac{1}{\sqrt{(3E_{s}^2 -k^2)(E_{s}^2-k_{z}^2)}} \left(\begin{array}{c} E_{s}^2-k_{z}^2 \\ E_{s}k_{x} e^{-i\phi}+k_{y}k_{z}e^{2i\phi} \\ E_{s} 
k_{y} e^{i\phi}+k_{x} k_{z} e^{-2i\phi} \end{array}\right),
\end{align}
where $E_s = s\hbar v_F |\mathbf{k}|$ is the energy associated to each eigenfunction.

We reproduce also the characteristic frequencies for the model in Eq.~\eqref{eqn:hamilt_3f} that determine the changes in the linear dependence of the optical conductivity, obtained previously in Ref.~\onlinecite{flicker_chiral_2018},
\begin{align}\label{eqn:3f_frequencies}
\dfrac{\hbar\omega_1}{\mu} &= \frac{\sqrt{3} \cos (\phi + \pi/6 )}{ \cos(\phi)}, &
\dfrac{\hbar\omega_2}{\mu} &= \frac{\sqrt{3} \cos (\phi + \pi/6 )}{ \cos(\phi - 2\pi/3)}, \nonumber \\
\dfrac{\hbar\omega_3}{\mu} &= \frac{\sqrt{3}\cos(-\phi+\pi/2)}{\cos(-\phi+\pi/3)}, &
\dfrac{\hbar\omega_4}{\mu}  &=\frac{\sqrt{3}\cos(\phi-\pi/6)}{\cos(\phi)}, \nonumber \\
\dfrac{\hbar\omega_5}{\mu}  &= \frac{\sqrt{3}\cos(-\phi+\pi/6)}{\cos(-\phi+\pi/3)}, &
\dfrac{\hbar\omega_6}{\mu} &=\frac{\sqrt{3}\sin \phi}{\cos(\phi-2\pi/3)}.
\end{align}
\subsection{Fourfold fermion}
For $a=3$, $b=-1$ ($\chi=\chi_0=\arctan(-1/3)$) we can write the Hamiltonian describing the fourfold fermion in Eq.~\eqref{eqn:ham_4fold} as $H_{4f}^{\chi_0}(\mathbf{k})=\hbar v_F \mathbf{k}\cdot \mathbf{S}_{3/2}$, where $\mathbf{S}_{3/2}=(S_{3/2,x},S_{3/2,y},S_{3/2,z})$ are three spin-3/2 matrices

\begin{align}
\label{eqn:spin1matrices}
S_{3/2,x}=&
\begin{pmatrix}
0 & 0 & 0 & \sqrt{3} \\
0 & 0 & \sqrt{3} & -2\\
0 & \sqrt{3} & 0 & 0 \\
\sqrt{3} & -2 & 0 & 0
\end{pmatrix},\nonumber \\
S_{3/2,y}=&
\begin{pmatrix}
0 & 0 & 0 & -\mi \sqrt{3} \\
0 & 0 & -\mi \sqrt{3} & -2 \mi\\
0 & \mi\sqrt{3} & 0 & 0 \\
\mi\sqrt{3} & 2\mi & 0 & 0
\end{pmatrix},\nonumber \\
S_{3/2,y}=&
\begin{pmatrix}
3 & 0 & 0 & 0 \\
0 & -1 & 0 &0\\
0 & 0 & -3 & 0 \\
0 & 0 & 0 & 1
\end{pmatrix},
\end{align}
with commutation relations $\left[S_{3/2,i},S_{3/2,j}\right]=2 \mi \epsilon_{ijk} S_{3/2,k}$.

For any value of $\chi$, the characteristic frequencies where the linear conductivity of the fourfold fermion changes slope were obtained in Ref.~\onlinecite{flicker_chiral_2018}.
Defining the momentum high symmetry directions $\mathbf{k}^{100} = k(1,0,0)$ and $\mathbf{k}^{111} = k(1,1,1)/\sqrt{3}$, the fourfold optical conductivity is determined by the following activation frequencies:
\begin{align}
\label{eqn:4f_frequencies}
\dfrac{\hbar\omega_1}{\mu} &= \frac{E_1(\mathbf{k}^{111})-E_2(\mathbf{k}^{111})}{E_1(\mathbf{k}^{111})}, & \dfrac{\hbar\omega_2}{\mu}  =  \frac{E_1(\mathbf{k}^{100})-E_2(\mathbf{k}^{100})}{E_1(\mathbf{k}^{100})}, 
\nonumber \\
\dfrac{\hbar\omega_3}{\mu}  &= \frac{E_1(\mathbf{k}^{100})-E_2(\mathbf{k}^{100})}{E_2(\mathbf{k}^{100})}, &  
\dfrac{\hbar\omega_4}{\mu}  = \frac{E_1(\mathbf{k}^{111})-E_2(\mathbf{k}^{111})}{E_2(\mathbf{k}^{111})}, 
\nonumber \\
\dfrac{\hbar\omega_5}{\mu}  &= \frac{E_1(\mathbf{k}^{100})-E_3(\mathbf{k}^{100})}{E_1(\mathbf{k}^{100})}, &  \dfrac{\hbar\omega_6}{\mu}  =  \frac{E_1(\mathbf{k}^{111})-E_3(\mathbf{k}^{111})}{E_1(\mathbf{k}^{111})}, &
\nonumber  \\
\dfrac{\hbar\omega_7}{\mu}  &= \frac{E_1(\mathbf{k}^{111})-E_4(\mathbf{k}^{111})}{E_1(\mathbf{k}^{111})},  & \dfrac{\hbar\omega_8}{\mu}  = \frac{E_2(\mathbf{k}^{111})-E_4(\mathbf{k}^{111})}{E_2(\mathbf{k}^{111})}, & 
\nonumber  \\
\dfrac{\hbar\omega_9}{\mu}  &= \frac{E_2(\mathbf{k}^{100})-E_4(\mathbf{k}^{100})}{E_2(\mathbf{k}^{100})}.   
\end{align}

\section{\label{sec:smoothing} Temperature and smoothing of the step function}

\begin{figure}[h!]
    \centering
    \includegraphics[width=\linewidth]{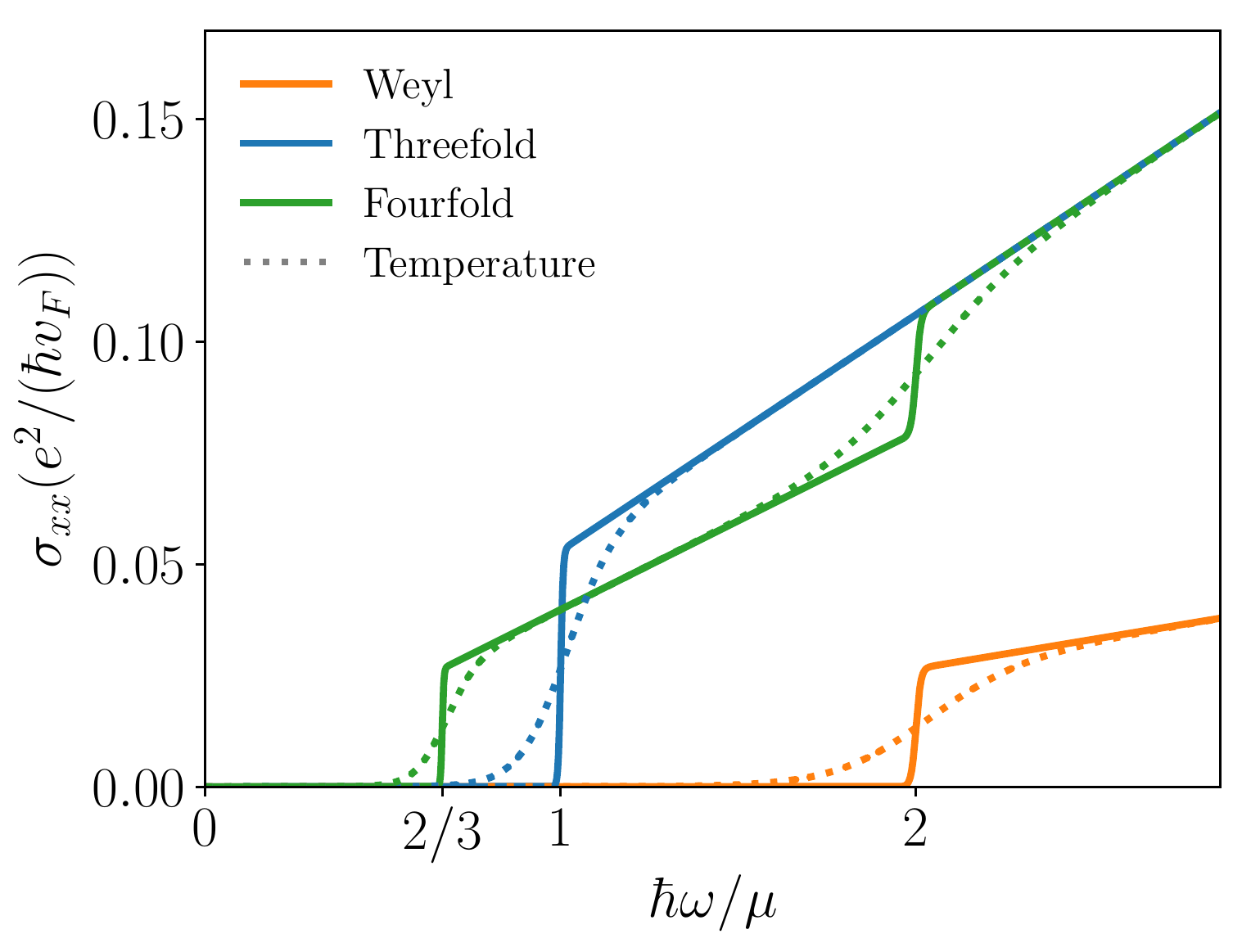}
    \caption{Optical conductivity of symmetric models with finite temperature. The solid lines correspond to that of Fig.~\ref{fig:Optcond_symmetric}  and the dashed lines (with the same color coding) are calculated with the exact analytic expressions Eqs.~\eqref{eqn:condweyl}, \eqref{eqn:optcond3ftemperature} and \eqref{eqn:cond4ftemperature} for finite temperature with $1/\beta=10^{-1}\mu$.}
    \label{fig:temperature}
\end{figure}
In Sec.~\ref{sec:lowenergymodels} we have derived analytic expressions for the optical conductivity of the symmetric threefold  and symmetric fourfold fermions, Eqs.~\eqref{eqn:optcond3ftemperature} and ~\eqref{eqn:cond4ftemperature} respectively, for any temperature $T=1/(k_B\beta)$. In Fig.~\ref{fig:temperature} we plot the optical conductivities for the twofold, symmetric threefold and symmetric fourfold fermions at zero temperature and at a finite (unrealistic) temperature $1/\beta=10^{-1}\mu$ to illustrate the smoothing of the step functions at the characteristic frequencies. In units of $\mu$ the broadening, set by $\mu\beta$, is larger for the step function at $2\mu$ than at $\mu$ or $2/3\mu$, which is clearly visible in Fig.~\ref{fig:temperature}.

The smoothing due to a finite temperature is visible as well in our calculations for realistic models in Sec.~\ref{sec:realistic}.

\begin{figure*}
    \centering
    \includegraphics[width=\linewidth]{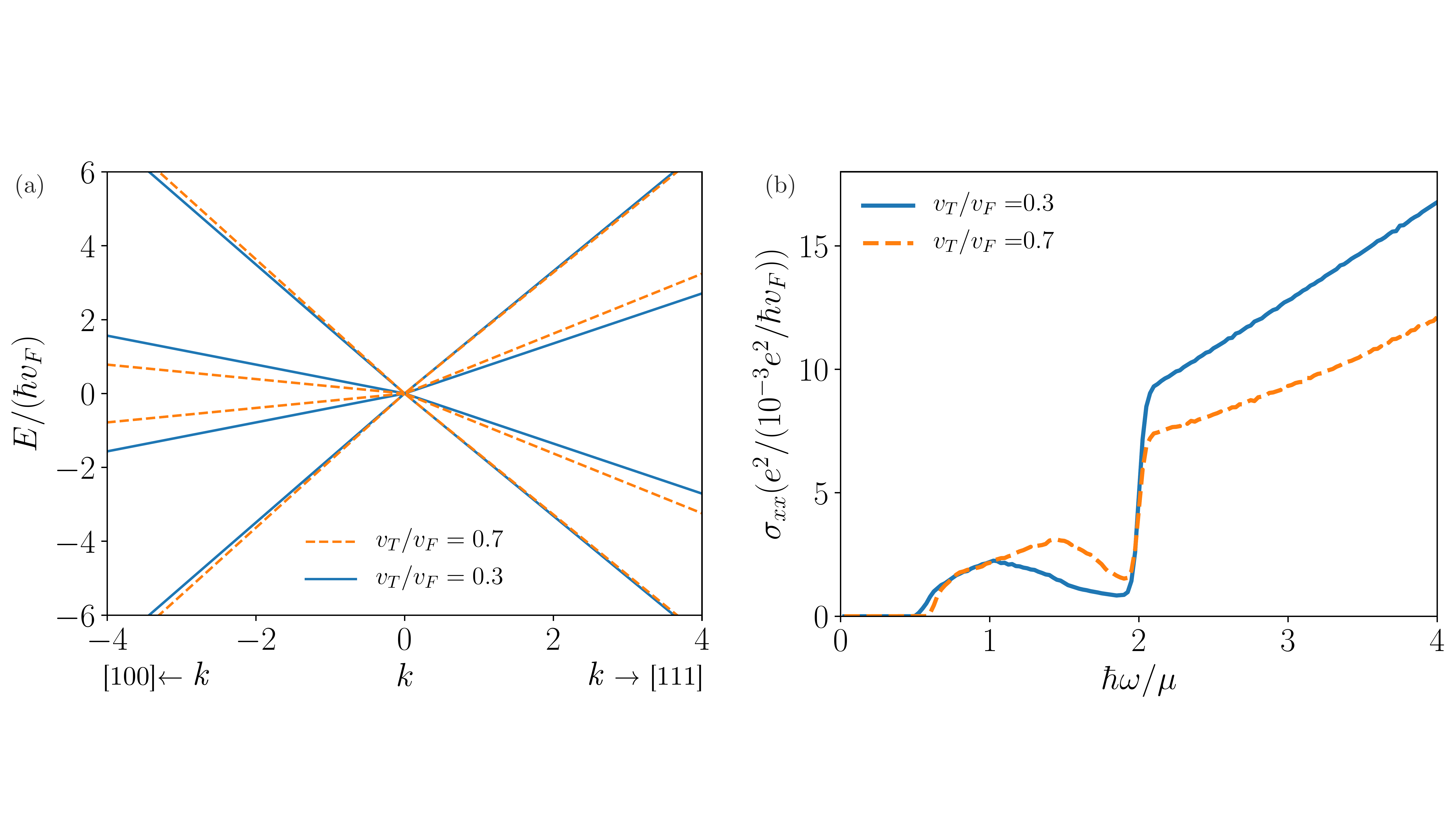}
    \caption{Optical conductivity of a tetrahedral fourfold fermion. (a) Band structure and (b) optical conductivity corresponding to the tetrahedral fourfold fermion with two different values of the ratio $v_{T}/v_F$ (see Eq.~
    \eqref{eq:tetrahedralH}).}
    \label{fig:tetrahedral}
\end{figure*}

\section{Tetrahedral fourfold\label{sec:4ftetra}}

The tetrahedral spin-$3/2$ fermions in space groups 195--198 arise upon breaking the fourfold rotational symmetry in space groups 207--214. At linear order, the Hamiltonian admits an extra term compared to the octahedral fourfold in space groups 195--198 and takes the form

\begin{equation}
\label{eq:tetrahedralH}
H_{4f,T}=H_{4f}+\hbar v_T\left(
\begin{array}{cccc}
0 & k_z & -\sqrt{3}k_x & ik_y \\
k_z & \frac{2k_z}{\sqrt{3}} & ik_y &\frac{k_x-2ik_y}{\sqrt{3}} \\
-\sqrt{3}k_x & -ik_y & 0 & -k_z \\
-ik_y & \frac{k_x+2ik_y}{\sqrt{3}} & -k_z & -\frac{2k_z}{\sqrt{3}}
\end{array}
\right),
\end{equation}
where $H_{4f}$ is the octahedral fourfold Hamiltonian given in Eq.~\eqref{eqn:ham_4fold}. The parameter $v_T$ is proportional to the strength of the fourfold rotational symmetry breaking. 

By changing $v_T$ we introduce a tilt in the bands (see Fig.~\eqref{fig:tetrahedral}~(a)), breaking the full rotational symmetry and leading to a different optical conductivity compared to Eq.~\eqref{eqn:cond4ftemperature}. The optical conductivity obtained for the tetrahedral fourfold fermion is shown in Fig.~\ref{fig:tetrahedral}~(b).

\section{Imaginary part of the optical conductivity $\sigma_{\Im}$ from Kramers-Kronig relations}
\label{sec:KK}
The optical conductivity is a complex quantity with real and imaginary parts $\sigma=\sigma_\Re+i\sigma_{\Im}$ which are related by the Kramers-Kronig relations~\cite{dresselhaus_optical_nodate}. In section \ref{sec:lowenergymodels} we have calculated the absorptive (real) part of the optical conductivity. Using the Kramers-Kronig relations we can obtain the dispersive (imaginary) part of the optical conductivity. The Kramers-Kronig relations are commonly written as 
\begin{align}
\label{eqn:KKfirst}
\sigma_{\Re}(\omega)=&\frac{1}{\pi}\mathcal{P}\int_{-\infty}^{\infty}\mathrm{d}x\frac{\sigma_{\Im}(x)}{x-\omega},\\
\sigma_{\Im}(\omega)=&-\frac{1}{\pi}\mathcal{P}\int_{-\infty}^{\infty}\mathrm{d}x\frac{\sigma_{\Re}(x)}{x-\omega},
\end{align}
where $\mathcal{P}$ denotes the Cauchy principal value. To calculate it we follow the procedure in Ref.~\onlinecite{dresselhaus_optical_nodate} and subtract the singularity at $\omega$ 
\begin{align}
    \sigma_{\Re}(\omega)+i\sigma_{\Im}(\omega)=\frac{1}{i\pi}\int_{-\infty}^{\infty}\mathrm{d}x\parent{\frac{\sigma(x)-\sigma(\omega)}{x-\omega}}\parent{\frac{x+\omega}{x+\omega}}.
\end{align}
Using now that the real part is even and the imaginary part is odd in frequencies we obtain
\begin{align}
    \label{eqn:KKreal}
    \sigma_{\Re}(\omega)=&\frac{2}{\pi}\int_{0}^{\infty}\mathrm{d}x\frac{x\sigma_{\Im}(x)-\omega\sigma_{\Im}(\omega)}{x^2-\omega^2},\\
    \sigma_{\Im}(\omega)=&-\frac{2\omega}{\pi}\int_0^{\infty}\mathrm{d}x\frac{\sigma_{\Re}(x)-\sigma_{\Re}(\omega)}{x^2-\omega^2}.
    \label{eqn:KKim}
\end{align}

Since the low-energy models that we used in section \ref{sec:lowenergymodels} to calculate the real part of the optical conductivity have unbounded linearly dispersing bands, we regularize the upper limit in the integrals in Eqs.~\eqref{eqn:KKreal} and \eqref{eqn:KKim} using a cutoff frequency $\Lambda$. As discussed in the main text, the real part of the optical conductivity of all chiral multifold fermions is a piecewise function of the form $\sigma_{\Re}(\omega)= \sum^{N-1}_{i=0}\sigma_i(\omega)=\sum_{i=0}^{N-1}S_i \omega \Theta(\omega_{i+1}-\omega_i)$.
The subindex $i$ is associated to each characteristic frequency $\omega_i$ where the slope of the optical conductivity changes (see Appendix \ref{app:symmetric}), where $\omega_0=0$ and $\omega_N=\Lambda/\hbar$ is the cutoff frequency, and $N$ is the number of different frequency regions. In particular, $N=7$ and $N=10$ for threefold and fourfold fermions as dictated by Eq.~\eqref{eqn:3f_frequencies} and \eqref{eqn:4f_frequencies} respectively. Using this partition for the optical conductivity we can rewrite now Eq.~(\ref{eqn:KKim}) as

\begin{widetext}
\begin{eqnarray}
\sigma_{\Im}\parent{\omega,\Lambda}
&=&-\frac{2\omega}{\pi}\left[\sum_{i=0}^{N-1}\int_{\omega_i}^{\omega_{i+1}}\frac{\sigma_i(x)}{x^2-\omega^2}\dd x-\int_{0}^{\infty}\frac{\sigma_\Re(\omega)}{x^2-\omega^2}\dd x \right]\\
&=&
-\frac{1}{\pi}\left[ \sigma_\Re(\omega)\log\left|\frac{\Lambda+\hbar\omega}{\Lambda-\hbar\omega}\right|+\omega\sum_{i=0}^{N-1}S_i\log\left| \frac{\omega_{i+1}^2-\omega^2}{\omega_i^2-\omega^2}  \right|\right],
\label{eqn:Imsigma}
\end{eqnarray}
This expression can be evaluated analytically for the cases of the twofold (Weyl), the symmetric threefold and the symmetric fourfold fermions presented in Table~\ref{table:optconds} in Eq.~\eqref{eqn:Imsigma}. 
For the Weyl fermion we obtain 
\begin{equation}
\label{eq:imweyl}
    \sigma_{\Im{},W}(\omega)=
    -\frac{\omega e^2}{24\pi^2\hbar v_F}
    \left[\log \left|
    \frac{\Lambda^2-(\hbar\omega)^2}{4\mu^2-(\hbar\omega)^2}\right|
    +\Theta(\hbar\omega-2\mu)\log\left|\frac{\Lambda+\hbar\omega}{\Lambda-\hbar\omega}\right|\right].
\end{equation}
We take the result obtained for the symmetric threefold in Eq.~\eqref{eqn:optcond3f}, and we obtain the corresponding imaginary part 
\begin{equation}
\label{eq:im3f}
    \sigma_{\Im,3f}^{\phi_0}(\omega)=
    -\frac{\omega e^2}{6\pi^2\hbar v_F}
    \left[\log \left|
    \frac{\Lambda^2-(\hbar\omega)^2}{\mu^2-(\hbar\omega)^2}\right|
    +\Theta(\hbar\omega-\mu)\log\left|\frac{\Lambda+\hbar\omega}{\Lambda-\hbar\omega}\right|\right].
\end{equation}
For the symmetric fourfold fermion
\begin{equation}
\label{eq:im4f}
    \sigma_{\Im,4f}^{\chi_0}(\omega)=
    -\frac{\omega e^2}{24\pi^2\hbar v_F}
    \left[4\log \left|
    \frac{\Lambda^2-(\hbar\omega)^2}{4\mu^2-(\hbar\omega)^2}\right|
    +
    3\log \left|
    \frac{36\mu^2-9(\hbar\omega)^2}{4\mu^2-9(\hbar\omega)^2}\right|
    +\left(3\Theta\parent{\hbar\omega-\frac{2\mu}{3}}+\Theta(\hbar\omega-2\mu)\right)\log\left|\frac{\Lambda+\hbar\omega}{\Lambda-\hbar\omega}\right|\right].
\end{equation}

For the non-symmetric multifold fermions, the characteristic frequencies can be calculated analytically for each $\phi,\chi$ using Eqs.~\eqref{eqn:3f_frequencies} and \eqref{eqn:4f_frequencies}. The slopes for each piece $S_i$ can be calculated numerically and introduced in Eq.~\eqref{eqn:Imsigma}.

\end{widetext}

\section{Sum rules}
\label{sec:sumrules}
\begin{figure}
    \centering
    \includegraphics[width=\linewidth]{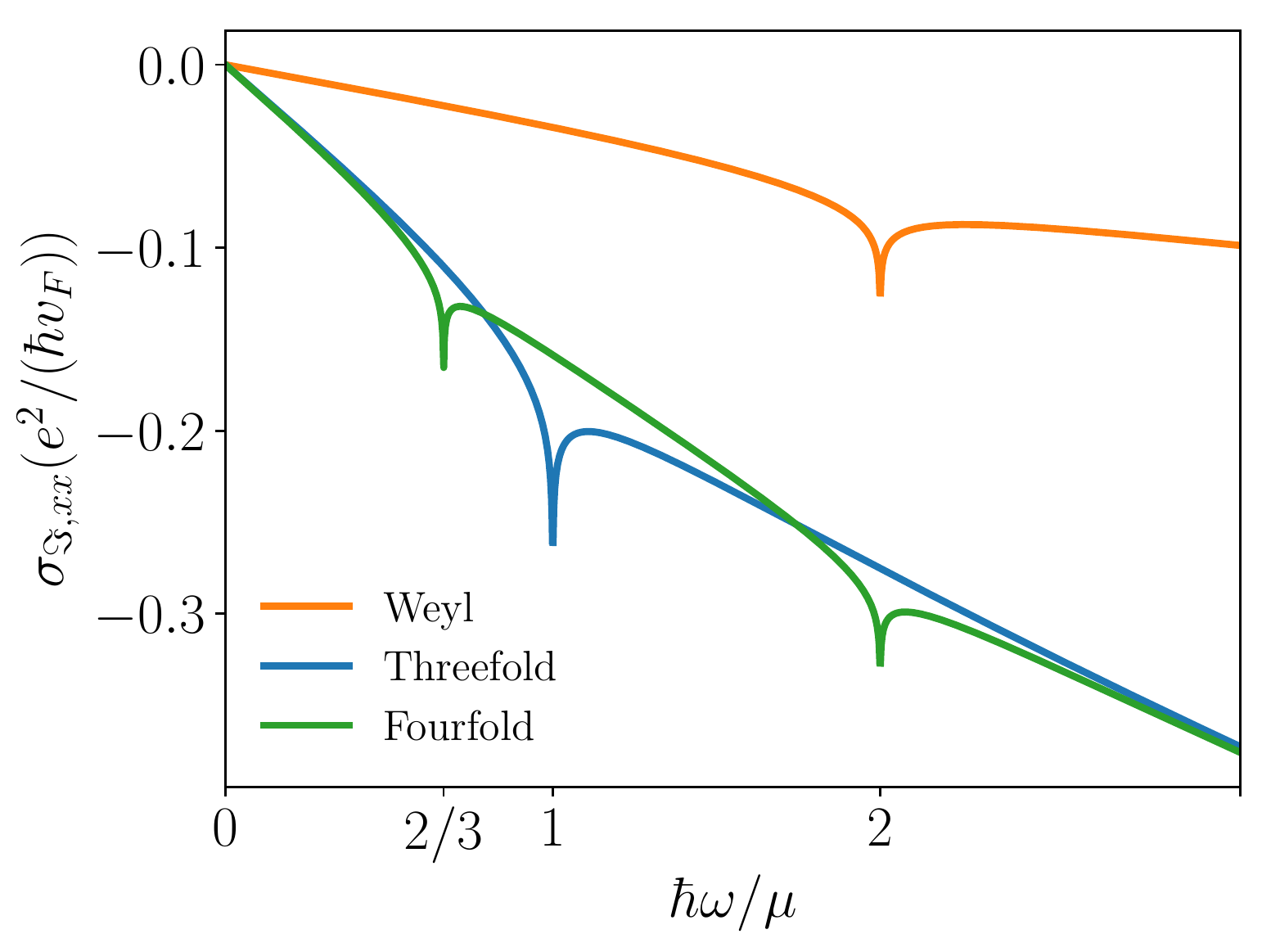}
    \caption{Imaginary part of the optical conductivity for a Weyl fermion (orange), a symmetric threefold fermion (blue), and a symmetric fourfold fermion (green) as dictated by Eqs.~\eqref{eq:imweyl}, \eqref{eq:im3f}, \eqref{eq:im4f} with $\Lambda=100\mu$.}
    \label{fig:my_label}
\end{figure}
Optical sum rules relate the real part of the optical conductivity with the total number of particles in the system, and are obtained as the integral of the optical conductivity to all frequencies,
\begin{eqnarray}
\label{eqn:sumrule}
\braket{\sigma}=\hbar^2\int_0^{\infty}\dd\omega\sigma_{\Re}(\omega).
\end{eqnarray}
As for the Kramers-Kronig relations, the unbounded linear dispersion of the effective low energy models requires us to insert a cutoff frequency $\Lambda$ in Eq.~\eqref{eqn:sumrule} to regularize the integral. As discussed in the previous section, we will use that the optical conductivity of these models is of the form $\sigma_{\Re}(\omega)= \sum^{N-1}_{i=0}\sigma_i=\sum_{i=0}^{N-1}S_i \omega \Theta(\omega_{i+1}-\omega_i)$ for both symmetric and non symmetric cases. Introducing this general form in Eq.~\eqref{eqn:sumrule} as well as the cut-off $\Lambda$ we obtain a general expression for the sum rule for all multifold fermions:
\begin{eqnarray}
\hbar^2\int_0^{\Lambda/\hbar}\dd\omega\sigma_{\Re}(\omega)&=
\nonumber
&\hbar^2\sum_{i=0}^{N-1}\int_{0}^{\Lambda/\hbar}\mathrm{d}\omega S_i \omega \Theta(\omega_{i+1}-\omega_i) \\
&=&\hbar^2\frac{e^2}{2\hbar v_F}\sum_{i=0}^{N-1}S_i(\omega_{i+1}^2-\omega_i^2).
\label{eqn:Sumrulemultifolds}
\end{eqnarray}
To obtain analytic results for the symmetric cases (see Sec.~\ref{subsec:lowenergysym}) we can insert the optical conductivities in Table~\ref{table:optconds} in Eq.~\eqref{eqn:Sumrulemultifolds}. In the twofold (Weyl) case we obtain
\begin{equation}
\braket{\sigma}_{2f}=\frac{e^2}{48\pi\hbar\vf}\parent{\Lambda^2-4\mu^2}.
\end{equation}
For the symmetric threefold fermion we obtain that
\begin{eqnarray}
\braket{\sigma}_{3f}=\frac{e^2}{12\pi\hbar\vf}\parent{\Lambda^2-\mu^2}.
\end{eqnarray}
In the symmetric fourfold case the optical sum rule is 
\begin{eqnarray}
\braket{\sigma}_{4f}=\frac{e^2}{12\pi\hbar v_F}\parent{\Lambda^2-\frac{4}{3}\mu^2}.
\end{eqnarray}

For the non symmetric cases the frequencies at which the linear dependence of the optical conductivity on $\omega$ changes are given by Eqs.~\eqref{eqn:3f_frequencies} and \eqref{eqn:4f_frequencies} for the threefold and fourfold fermions, respectively. In this case, the linear dependence $S_i$ in each section $\omega_i<\omega<\omega_{i+1}$ can be computed numerically and substituted in Eq.~\eqref{eqn:Sumrulemultifolds} to obtain the corresponding sum rule.

Finally, note that the Drude peak will contribute to the sum rule as well. Extending the results of Ref.~\onlinecite{sabio_f-sum_2008} to three-dimensions, we expect its contribution to be proportional to $\mu^2$.

\section{Tight-binding models and orbital embedding}
\label{sec:orbitalembedding}

In Sec. \ref{sec:realistic} we have calculated the optical conductivity of materials described by tight-binding models in space groups 199 and 198. The tight-binding model for SG198 and a detailed discussion on its construction without orbital embedding can be found in Ref.~\onlinecite{chang_unconventional_2017}. The inclusion of the orbital embedding for SG198, together with the construction of the tight-binding model for SG199 is discussed in Ref.~\onlinecite{flicker_chiral_2018}. For convenience we revisit here how to include the orbital embedding for the models we used in the main text. 

Materials in space group 199 have body-centered cubic structures with Bravais lattice vectors 
\begin{align}
R_1&=\frac{a}{2}(-\mathbf{\hat{x}}+\mathbf{\hat{y}}+\mathbf{\hat{z}}),\nonumber\\
R_2&=\frac{a}{2}(\mathbf{\hat{x}}-\mathbf{\hat{y}}+\mathbf{\hat{z}}),\nonumber\\
R_3&=\frac{a}{2}(\mathbf{\hat{x}}+\mathbf{\hat{y}}-\mathbf{\hat{z}}).
\end{align}
To construct the tight-binding model considering the symmetries of SG199 we place spinless $s$-orbitals in the positions $\mathbf{q}_i$, given by
\begin{align}\label{eqn:Wyckoff199}
\mathbf{q}_1 &= (u,u,u), \nonumber\\
\mathbf{q}_2 &= (\frac{1}{2}-u,\frac{1}{2},0), \nonumber \\
\mathbf{q}_3 &= (0,\frac{1}{2}-u,\frac{1}{2}),\nonumber \\
\mathbf{q}_4  &=(\frac{1}{2},0,\frac{1}{2}-u),
\end{align}
where $|u|< 1/2$, and $\mathbf{q}_i$ is expressed in reduced coordinates, \emph{i.e.}, in units of $R_i$.

Then, we can write the tight-binding Hamiltonian used in Sec.~\ref{subsec:SG199} for a material in SG199 as $H_{199}(u,\mathbf{k})=V^{\dagger}(u,\mathbf{k})H_0(\mathbf{k})V(u,\mathbf{k})$, where

\begin{equation}
H_0(\mathbf{k})=\left(\begin{array}{cccc}
0 & 1 & 1 & 1 \\
1 & 0 & e^{-i\mathbf{k}\cdot\mathbf{R}_3} & e^{i\mathbf{k}\cdot\mathbf{R}_2} \\
1 & e^{i\mathbf{k}\cdot\mathbf{R}_3} & 0 & e^{-i\mathbf{k}\cdot\mathbf{R}_1}\\
1 & e^{-i\mathbf{k}\cdot\mathbf{R}_2} & e^{i\mathbf{k}\cdot\mathbf{R}_1} & 0
\end{array}\right),
\end{equation}
and

\begin{widetext}
\begin{equation}
V(u,\mathbf{k})=\left(\begin{array}{cccc}
e^{i\mathbf{k}\cdot\mathbf{q}_1} & 0 & 0 & 0 \\
0 & e^{i\mathbf{k}\cdot\mathbf{q}_2} & 0 & 0 \\
0 & 0 & e^{i\mathbf{k}\cdot\mathbf{q}_3} & 0 \\
0 & 0 & 0 & e^{i\mathbf{k}\cdot\mathbf{q}_4}
\end{array}
\right).
\end{equation}

For SG198 the tight-binding Hamiltonian presented in Ref.~\onlinecite{chang_unconventional_2017} was modified in Ref.~\onlinecite{flicker_chiral_2018} to take into account the orbital embedding. In the original tight-binding Hamiltonian\cite{chang_unconventional_2017} $\mathcal{H}(\mathbf{k})$ the atoms are located in the positions 

\begin{align}
\mathbf{q}_A=\left(0,0,0\right), \quad
\mathbf{q}_B=\left(\frac{1}{2},\frac{1}{2},0\right), \quad
\mathbf{q}_C=\left(\frac{1}{2},0,\frac{1}{2}\right), \quad
\mathbf{q}_D=\left(0,\frac{1}{2},\frac{1}{2}\right),
\end{align}
given in reduced coordinates. To take into account the orbital embedding, the new atomic positions 
\begin{align}
\mathbf{q}_A=\left(x,x,x\right), \quad
\mathbf{q}_B=\left(\frac{1}{2}+x,\frac{1}{2}-x,-x\right), \quad
\mathbf{q}_C=\left(\frac{1}{2}-x,-x,\frac{1}{2}+x\right), \quad
\mathbf{q}_D=\left(-x,\frac{1}{2}+x,\frac{1}{2}-x\right),
\end{align}
were introduced in Ref.~\onlinecite{flicker_chiral_2018} with $x=0.3959$ for RhSi, according to their ab-initio calculations. In Sec.~\ref{subsec:SG198} we have calculated the optical conductivity of RhSi using the tight-binding Hamiltonian $H_{198}(x,\mathbf{k})=U_x(\mathbf{k})^{\dagger}\mathcal{H}(\mathbf{k})U_x(\mathbf{k})$ with
\begin{equation}
U_x(\mathbf{k})=\exp\left[\left(
\begin{array}{cccc}
ix(k_1+k_2+k_3) & 0 & 0 & 0 \\
0 & ix(k_1-k_2-k_3) & 0 & 0 \\
0 & 0 & ix(k_3-k_2-k_1) & 0 \\
0 & 0 & 0 & ix(k_2-k_1-k_3)
\end{array}
\right)\right],
\end{equation}
and $\mathcal{H}(\mathbf{k})$ the tight-binding Hamiltonian without spin-orbit coupling presented in Ref.~\onlinecite{chang_unconventional_2017}, which reads
\begin{eqnarray}
\label{eqn:tb198}
\mathcal{H}(\mathbf{k}) &=& v_{1}\bigg[\tau^{x}\mu^{0}\cos\left(\frac{k_{x}}{2}\right)\cos\left(\frac{k_{y}}{2}\right) + \tau^{x}\mu^{x}\cos\left(\frac{k_{y}}{2}\right)\cos\left(\frac{k_{z}}{2}\right) + \tau^{0}\mu^{x}\cos\left(\frac{k_{z}}{2}\right)\cos\left(\frac{k_{x}}{2}\right)\bigg] \nonumber \\
&+&  v_{p}\bigg[\tau^{y}\mu^{z}\cos\left(\frac{k_{x}}{2}\right)\sin\left(\frac{k_{y}}{2}\right) + \tau^{y}\mu^{x}\cos\left(\frac{k_{y}}{2}\right)\sin\left(\frac{k_{z}}{2}\right) + \tau^{0}\mu^{y}\cos\left(\frac{k_{z}}{2}\right)\sin\left(\frac{k_{x}}{2}\right)\bigg] \nonumber \\
&+&v_{2}\bigg[\cos\left(k_{x}\right) + \cos\left(k_{y}\right) + \cos\left(k_{z}\right)\bigg]\tau^{0}\mu^{0},
\end{eqnarray}
where $\tau^{i}$ and $\mu^{i}$, $i=x,y,z$, are the three Pauli matrices for spin-1/2, $\tau^{0}=\mu^{0}=\mathbb{1}$ is the $2\times2$ identity matrix, and $\tau^{i}\mu^{j}\equiv\tau^{i}\otimes\mu^{j}$ is a short-hand notation for the Kronecker product. For RhSi the values of the tight-binding parameters are $v_1=0.55$, $v_2=0.16$, and $v_p=-0.76$, obtained in Ref.~\onlinecite{chang_unconventional_2017} by fitting the bands of the tight-binding Hamiltonian in Eq.~\eqref{eqn:tb198} to their first-principles calculations.

\end{widetext}

\bibliographystyle{apsrev4-1}

%

\end{document}